%% file: paper.tex
\documentclass[runningheads]{llncs}

\usepackage{amssymb}
\usepackage{amsmath}
\usepackage{graphicx}
\usepackage{url}
\usepackage{titleref}
\usepackage{picins}
\usepackage{subfigure}
\usepackage{lscape}
\usepackage{enumerate}
\usepackage{longtable}
\usepackage{wrapfig}
\usepackage{float}
\usepackage{multicol}
\usepackage{multirow}
\usepackage{booktabs}
\usepackage{microtype}
\usepackage[ansinew]{inputenc} %Windows
\usepackage[pdftex]{color}
\usepackage{cancel}
\usepackage{paralist}
\usepackage[pdfsubject={Business Process Management},
							pdftitle={Maximal Structuring of Acyclic Process Models},
							pdfauthor={Artem Polyvyanyy, Luciano Garcia-Banuelos, Dirk Fahland, and Mathias Weske},
							pdfkeywords={Process modeling, structured process model, maximal structuring, model transformation, fully concurrent bisimulation},
							colorlinks=false,
							pdfborder={0 0 0}]{hyperref}
\usepackage[algoruled,noend]{algorithm2e}
\usepackage{savesym}
\savesymbol{vec}
\usepackage{MnSymbol}
\hypersetup{bookmarksdepth=3}

\graphicspath{{}}

\input{artem.sty}
\begin{document}
%%%%%%%%%%%%%%%%%%%%%%%%%%%%%%%%%%%%%%%%%%%%%%%%%%%%%%%%%%%%%%%%%%%%%%%%%%%%%%%%%%%%%%%%%%

\mainmatter

\title{Maximal Structuring of Acyclic Process Models}

\author{Artem Polyvyanyy\small{\inst{1}} \and Luciano Garc\'ia-Ba\~nuelos\small{\inst{2}} \and Dirk Fahland\small{\inst{3}} \and Mathias Weske\small{\inst{1}}}
\authorrunning{Polyvyanyy, A. \and Garc\'ia-Ba\~nuelos, L. \and Fahland, D. \and and Weske, M.}

\institute{
%Business Process Technology Group\\
Hasso Plattner Institute at the University of Potsdam, Germany\\
%Prof.-Dr.-Helmert-Str. 2--3, D-14482 Potsdam, Germany\\
\href{mailto:Artem.Polyvyanyy@hpi.uni-potsdam.de;Mathias.Weske@hpi.uni-potsdam.de}{\texttt{\{Artem.Polyvyanyy,Mathias.Weske\}@hpi.uni-potsdam.de}}
\and
Institute of Computer Science, University of Tartu, Estonia\\
%J. Liivi 2, Tartu 50409, Estonia\\
\href{mailto:Luciano.Garcia@ut.ee}{\texttt{Luciano.Garcia@ut.ee}}\\
\and
Eindhoven University of Technology, The Netherlands\\
\href{mailto:D.Fahland@tue.nl}{\texttt{D.Fahland@tue.nl}}\\
}

\maketitle

%%%%%%%%%%%%%%%%%%%%%%%%%%%%%%%%%%%%%%%%%%%%%%%%%%%%%%%%%%%%%%%%%%%%%%%%%%%%%%%%%%%%%%%%%%
\vspace{-6mm}
\input{abstract}
\vspace{-8mm}
%%%%%%%%%%%%%%%%%%%%%%%%%%%%%%%%%%%%%%%%%%%%%%%%%%%%%%%%%%%%%%%%%%%%%%%%%%%%%%%%%%%%%%%%%%

\input{introduction}
\input{preliminaries}
\input{structuring}
\input{maximal}
\input{related}
\input{conclusion}

\vspace{-2mm}
\bibliographystyle{splncs}
\bibliography{bibliography}

%%%%%%%%%%%%%%%%%%%%%%%%%%%%%%%%%%%%%%%%%%%%%%%%%%%%%%%%%%%%%%%%%%%%%%%%%%%%%%%%%%%%%%%%%%
\end{document}
%%%%%%%%%%%%%%%%%%%%%%%%%%%%%%%%%%%%%%%%%%%%%%%%%%%%%%%%%%%%%%%%%%%%%%%%%%%%%%%%%%%%%%%%%%

%% file: abstract.tex
\begin{abstract}
This paper contributes to the solution of the problem of transforming a process model with an arbitrary topology into an equivalent structured process model. 
In particular, this paper addresses the subclass of process models that have no equivalent well-structured representation but which, nevertheless, can be partially structured into their maximally-structured representation. 
The structuring is performed under a behavioral equivalence notion that preserves observed concurrency of tasks in equivalent process models. 
The paper gives a full characterization of the subclass of acyclic process models that have no equivalent well-structured representation but do have an equivalent maximally-structured one, as well as proposes a complete structuring method.
\end{abstract}

%% file: introduction.tex
%%%%%%%%%%%%%%%%%%%%%%%%%%%%%%%%%%%%%%%%%%%%%%%%%%%%%%%%%%%%%%%%%%%%%%%%%%%%%%%%%%%%%%%%%%
\section{Introduction}
\label{sec:introduction}
\vspace{-2mm}
%%%%%%%%%%%%%%%%%%%%%%%%%%%%%%%%%%%%%%%%%%%%%%%%%%%%%%%%%%%%%%%%%%%%%%%%%%%%%%%%%%%%%%%%%%

\enlargethispage{\baselineskip}

%[Intro]
%One can fairly adapt the ideas of Donald E. Knuth~\cite{Knuth74} to conclude that process modeling is both a science and an art. Process modeling does have an aesthetic sense. Just like when composing an opera or writing a novel, process modeling is carried out by humans who undergo creative practices when engineering a process model. Therefore, the very same process can be modeled in a myriad number of different ways. Once formalized, the process model can be analyzed by employing scientific methods.

\begin{figure}[t]
\vspace{-3mm}
\begin{center}
  \subfigure[]{\label{fig:and:prim:unstr}\includegraphics[scale = .55]{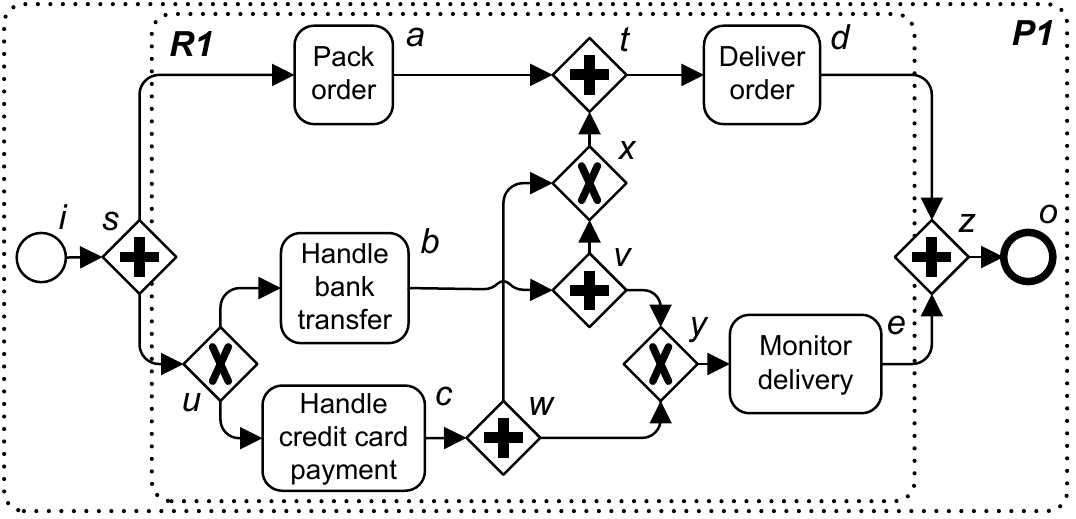}}
  \hspace{4mm}
  \subfigure[]{\label{fig:and:prim:str}\includegraphics[scale = .55]{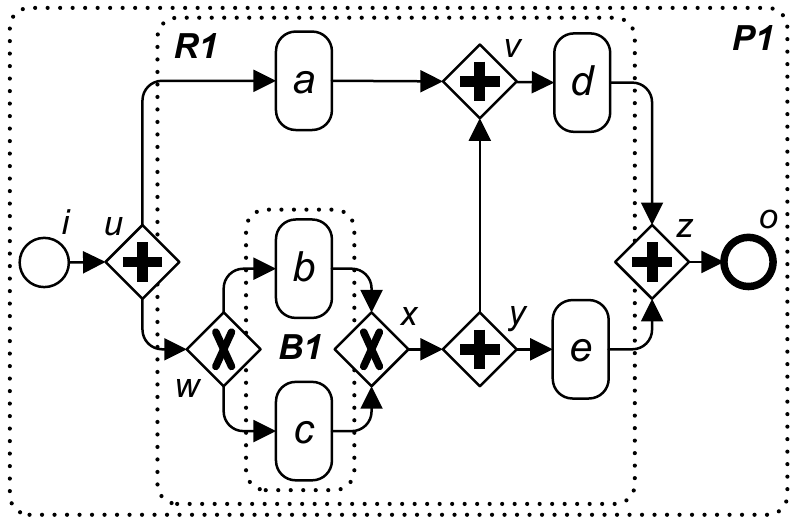}}
\end{center}
\vspace{-7mm}
\caption{\small{(a) Unstructured model, and (b) its equivalent maximally-structured version}}
\label{fig:and:prim}
\vspace{-6mm}
\end{figure}

Process models are usually represented as graphs, where nodes stand for tasks or decisions, and edges encode causal dependencies between adjacent nodes.
Common process modeling notations, such as Business Process Model and Notation (BPMN) or Event-driven Process Chains (EPC), allow process models to have almost any topology.
Structural freedom allows for a large degree of creativity when modeling.
Nevertheless, it is often preferable that models follow certain structural patterns.
%[Structural properties]
A well-known property of process models is that of (well-) structuredness~\cite{KHB00}.
A model is \emph{well-structured}, if for every node with multiple outgoing arcs (a split) there is a corresponding node with multiple incoming arcs (a join), such that the fragment of the model between the split and the join forms a single-entry-single-exit (SESE) process component; otherwise the model is unstructured.
\figurename~\ref{fig:and:prim:unstr} shows a process model.
Each dotted box defines a component composed from the arcs that are inside or intersect the box.
Split $s$ has corresponding join $z$; together they define SESE component $R1$.
Yet, split $u$ has no corresponding join and, thus, the model in \figurename~\ref{fig:and:prim:unstr} is unstructured.

%[Example]

The motivations for well-structured process modeling are manifold. Structured
models are easier to layout, understand, support, and analyze~\cite{PGD10}.
Consequently, some process modeling languages urge for structured modeling, \eg Business Process Execution Language (BPEL) and ADEPT.
We advocate for a different philosophy:
The modeling language should provide process modelers with a maximal degree of structural freedom to describe processes.
Afterwards, scientific methods can suggest (whenever possible) alternative formalizations that are better structured, preferably well-structured.

In previous work~\cite{PGD10}, we proposed a technique to automatically transform acyclic process models with arbitrary topologies into equivalent well-structured models.
The structuring is accomplished under a strong notion of behavioral equivalence called \emph{fully concurrent bisimulation}~\cite{KHB00,PGD10}.
As an outcome, the resulting well-structured models describe the same share of concurrency as the original unstructured models.
It was shown in~\cite{KHB00} (by means of a single example) and confirmed in~\cite{PGD10} (for the general case of acyclic models) that there exist process models that do not have an equivalent well-structured representation.
\figurename~\ref{fig:and:prim:unstr} is an example of such a model.
Though not completely structurable, this model can be partially structured to result in its maximally-structured version shown in \figurename~\ref{fig:and:prim:str}.
A process model is \emph{maximally-structured}, if every model that is equivalent with it has at least the same number of SESE components defined by pairs of a split and join node as the model itself.
Note that \figurename~\ref{fig:and:prim:str} uses short-names for tasks $(a,b,c \ldots )$, which appear next to each task in \figurename~\ref{fig:and:prim:unstr}.

After the initial investigations in~\cite{ElligerPW10}, this paper gives for the first time a complete solution to the problem of maximal structuring of acyclic process models.
We characterize the class of acyclic process models which do not have an equivalent well-structured representation, but which can, nevertheless, be maximally structured; and we provide a complete structuring method.

%[Structure]
The remainder of this paper proceeds as follows:
The next section gives preliminary definitions.
\sectionname~\ref{sec:structuring} discusses the structuring technique proposed in~\cite{PGD10}.
The technique is summarized as a chain of transformations.
We define for the first time the notion of a proper complete prefix unfolding which was sketched in~\cite{PGD10} and which is essential for obtaining sufficient behavioral information to allow maximal structuring.
\sectionname~\ref{sec:maximal:structuring} devises an extension of the structuring technique for maximal structuring of process models that do not
have an equivalent well-structured representation.
\sectionname~\ref{sec:conclusion} discusses related work and draws conclusions.

%%%%%%%%%%%%%%%%%%%%%%%%%%%%%%%%%%%%%%%%%%%%%%%%%%%%%%%%%%%%%%%%%%%%%%%%%%%%%%%%%%%%%%%%%%

%% file: preliminaries.tex
%%%%%%%%%%%%%%%%%%%%%%%%%%%%%%%%%%%%%%%%%%%%%%%%%%%%%%%%%%%%%%%%%%%%%%%%%%%%%%%%%%%%%%%%%%
\vspace{-3mm}
\section{Preliminaries}
\label{sec:preliminaries}
\vspace{-2mm}
%%%%%%%%%%%%%%%%%%%%%%%%%%%%%%%%%%%%%%%%%%%%%%%%%%%%%%%%%%%%%%%%%%%%%%%%%%%%%%%%%%%%%%%%%%

\enlargethispage{\baselineskip}

Preliminaries describe formalisms that will be used later to convey the findings.

%%%%%%%%%%%%%%%%%%%%%%%%%%%%%%%%%%%%%%%%%%%%%%%%%%%%%%%%%%%%%%%%%%%%%%%%%%%%%%%%%%%%%%%%%%
\vspace{-4mm}
\subsection{Process Models and Nets}
\label{subsec:pm:and:nets}
\vspace{-1mm}
%%%%%%%%%%%%%%%%%%%%%%%%%%%%%%%%%%%%%%%%%%%%%%%%%%%%%%%%%%%%%%%%%%%%%%%%%%%%%%%%%%%%%%%%%%

%[Intro]
This section introduces all subsequently required notions on process models.

%[Process model]
\vspace{-1mm}
\begin{definition}[Process model]\emph{
\label{def:process:model}
A \emph{process model} $P=(A,G,C,type,\mathcal{A},\mu)$ has a non-empty set $A$ of \emph{tasks}, a set $G$ of \emph{gateways}, $A \cap G = \emptyset$, and a set $C \subseteq (A \cup G) \times (A \cup G)$ of \emph{control flow} arcs of $P$;
$\mathit{type} : G \rightarrow \{\mathit{xor},\mathit{and}\}$ assigns to each gateway a type; $\mu : A \rightarrow \mathcal{A}$ assigns to each task a \emph{name} from $\mathcal{A} \neq \emptyset$.}
\end{definition}
\vspace{-1mm}

%[Details on process model]
\noindent
$A \cup G$ are the \emph{nodes} of $P$; a node $x\in{A \cup G}$ is a \emph{source} (\emph{sink}), \ifaof $\pre{x} = \emptyset$ ($\post{x} = \emptyset$), where $\pre{x}$ ($\post{x}$) stands for the set of immediate predecessors (successors) of $x$. We assume $P$ to have a single source and a single sink task; every node of $P$ is on a path from source to sink. Each task $a\in{A}$ has at most one incoming and at most one outgoing arc, \ie $\left| \pre{a} \right| \leq 1 \wedge \left| \post{a} \right| \leq 1$. Each gateway $g \in G$ is either a \emph{split} ($\left| \pre{g} \right| = 1 \wedge \left| \post{g} \right| > 1$) or a \emph{join} ($\left| \pre{g} \right| > 1 \wedge \left| \post{g} \right| = 1$).
The semantics of process models is usually defined by a mapping to Petri nets.

\vspace{-1.5mm}
\begin{definition}[Petri net]\emph{
\label{def:pn} A \emph{Petri net}, or a \emph{net}, $N = (P,T,F)$ has finite
disjoint sets $P$ of \emph{places} and $T$ of \emph{transitions}, and the
\emph{flow} relation $F \subseteq (P \times T) \cup (T \times P)$.
A \emph{net system} $(N,M)$ is a net $N$ with a \emph{marking} $M : P \to \mathbb{N}_0$ assigning each $p \in P$ a number $M(p)$ of \emph{tokens} in place $p$; $M_0$ denotes the \emph{initial} marking.
}\end{definition}
\vspace{-1.5mm}

\noindent
For a node $x \in P \cup T$, $\pre{x}=\{y \ | \ (y,x) \in F\}$ is a \emph{preset}, whereas $\post{x}=\{y \ | \ (x,y) \in F\}$ is a \emph{postset} of $x$; $\mathit{Min}(N)$ denotes the set of nodes of $N$ with an empty preset. For $X \subseteq P \cup T$, let $\pre{X}=\bigcup_{x \in X}\pre{x}$ and $\post{X}=\bigcup_{x \in X}\post{x}$. For a binary relation $R$ (\eg $F$ or $C$), $R^+$ and $R^\ast$ denote irreflexive and reflexive transitive closures of $R$.

%We identify $F$ with its characteristic function on the set $(P \times T) \cup (T \times P)$.
%A node $x\in{P \cup T}$ is an \emph{input} (\emph{output}) node of a node $y\in{P \cup T}$, \ifaof $x\in{\pre{y}}$ ($x\in{\post{y}}$).

A net $N = (P,T,F)$ is \emph{free-choice}, \ifaof $\forall \ p\in P, |\post{p}|>1 : \pre{(\post{p})}=\{p\}$. A \emph{labeled} net $N = (P,T,F,\mathcal{T},\lambda)$ has a function $\lambda : P \cup T \to \mathcal{T}$ that assigns each node a \emph{label} from $\mathcal{T}$, $\tau \in \mathcal{T}$. If $\lambda(t) \neq \tau$, then $t\in T$ is \emph{observable}; otherwise, $t$ is \emph{silent}.

\begin{wrapfigure}{r}{.5\textwidth}
\vspace{-11mm}
\begin{center}
  \includegraphics[scale = .46, trim = 2mm 0 0 0]{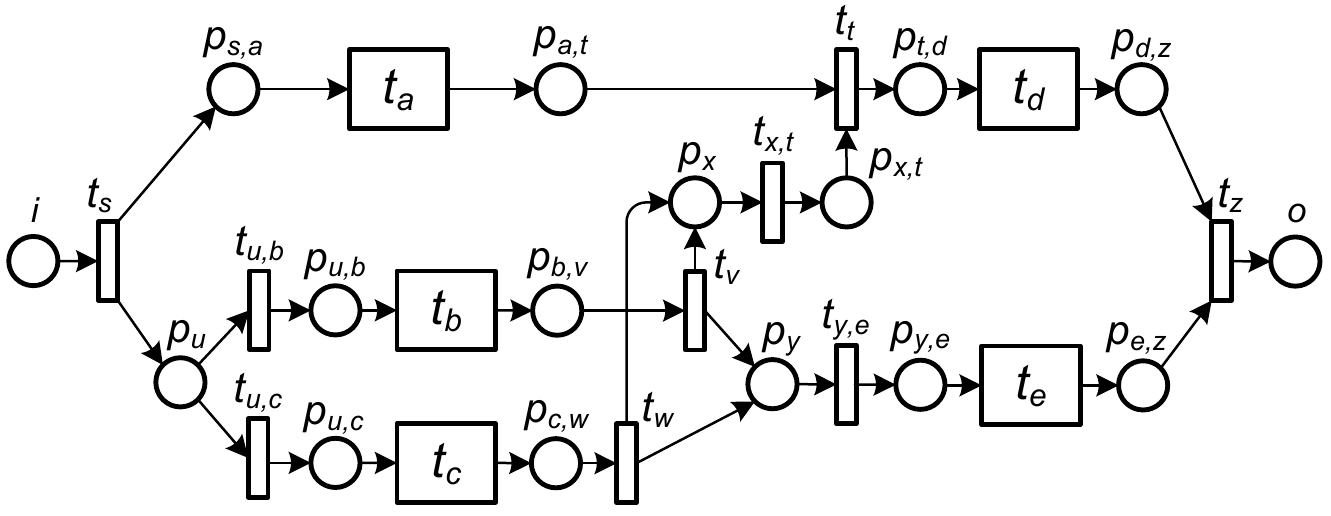}
\end{center}
\vspace{-6mm} 	\caption{\small{A WF-net that corresponds to the process
model in \figurename~\ref{fig:and:prim:unstr}}} 	\label{fig:net}
\vspace{-7mm}
\end{wrapfigure}
Every process model (Def.~\ref{def:process:model}) can be mapped to a
labeled free-choice net with a special structure, called WF-net~\cite{KiepuszewskiHA03,PGD10}; the net in
\figurename~\ref{fig:net} corresponds to the model in \figurename~\ref{fig:and:prim:unstr}. The execution semantics of the net (in terms of its token game) defines the semantics of the model.
In our work, we require process models to be sound~\cite{A97}, with the intuition that a model is sound, \ifaof its corresponding WF-system is sound.

\vspace{-3mm}
\subsection{Unfoldings}
\label{subsec:unfolding}
\vspace{-1mm}
%%%%%%%%%%%%%%%%%%%%%%%%%%%%%%%%%%%%%%%%%%%%%%%%%%%%%%%%%%%%%%%%%%%%%%%%%%%%%%%%%%%%%%%%%%

\enlargethispage{\baselineskip}

%[Intro]
%An unfolding of a system is another net (possibly infinite, but with a tree-like structure), such that whenever there is a conflict in the system, the unfolding splits into several independent copies of itself, one for each resolution of the conflict~\cite{NPW80,Engelfriet91,EspHel08}.
An unfolding of a net system is another net that explicitly represents all runs of the system in a possibly infinite, tree-like structure~\cite{NPW80,Engelfriet91,EspHel08}. In~\cite{McMillan95}, McMillan proposed an algorithm for the construction of a \emph{finite} initial part of the unfolding, which contains full information about the reachable states of a system -- a \emph{complete prefix unfolding}. Next, we present main notions of the theory of unfoldings. First, we define ordering relations between pairs of nodes in a net.

%[Def: Ordering relations]
\vspace{-1.5mm}
\begin{definition}[Ordering relations]\emph{
\label{def:order:rel} Let $N=(P,T,F)$ be a net, $x,y \in P \cup T$.
\begin{compactitem}
  \item $x$ and $y$ are in \emph{causal} relation, written $x \leadsto_N
y$, \ifaof $(x,y) \in F^+$. $y$ and $x$ are in \emph{inverse causal}
relation, written $y \leftrsquigarrow_N x$, \ifaof $x \leadsto_N
y$.
  \item $x$ and $y$ are in \emph{conflict}, $x \ \#_N \
y$, \ifaof there exist distinct transitions $t_1,t_2 \in T$, s.t.
$\pre{t_1} \cap \pre{t_2} \neq \emptyset$, and $(t_1,x),(t_2,y) \in
F^\ast$. If $x \ \#_N \ x$, then $x$ is in \emph{self-conflict}.
  \item $x$ and $y$ are \emph{concurrent}, $x \ ||_N \ y$, \ifaof
neither $x \leadsto_N y$, nor $y \leadsto_N x$, nor $x \ \#_N \ y$.
\end{compactitem}
The set $\mathcal{R}_N=\{\leadsto_N, \leftrsquigarrow_N, \#_N, ||_N\}$ forms
the \emph{ordering relations} of $N$. }\end{definition}
\vspace{-1.5mm}

\noindent
Note that in the following we omit subscripts of ordering relations where the context is clear.
A structure of an unfolding is given by an \emph{occurrence} net.

%[Def: Occurrence net]
\vspace{-1.5mm}
\begin{definition}[Occurrence net]\emph{
\label{def:occ:net} A net $N=(B,E,G)$ is an \emph{occurrence} net, \ifaof:
for all $b \in B$ holds $|\pre{b}| \leq 1$, $N$ is acyclic, for each $x \in
B \cup E$ the set $\{y \in B \cup E \ | \ (y,x) \in G^+ \}$ is finite, and no $e \in E$ is
in self-conflict. }\end{definition}
\vspace{-1.5mm}

\noindent The elements of $B$ and $E$ are called \emph{conditions} and
\emph{events}, respectively. Any two nodes of an occurrence net are either in
causal, inverse causal, conflict, or concurrency relation~\cite{NPW80}. An
unfolding of a system is closely related to the concept of a branching process
of a system. A branching process is an occurrence net where each node is
mapped to a node of the system.

%Let $N_1=(P_1,T_1,F_1)$ and $N_2=(P_2,T_2,F_2)$ be nets. A \emph{homomorphism} from $N_1$ to $N_2$ is a mapping $h : X_1 \rightarrow X_2$, such that: $h(P_1) \subseteq P_2$ and $h(T_1) \subseteq T_2$, and for all $t \in T_1$, the restriction of $h$ to $\pre{t}$ is a bijection between $\pre{t}$ in $N_1$ and $\pre{h(t)}$ in $N_2$; correspondingly for $\post{t}$ and $\post{h(t)}$.
%Homomorphism is a mapping that preserves the nature of nodes and the environment of transitions.
%the set of minimal nodes of $N$ with respect to the transitive closure of the flow relation, \ie the set of nodes that has an empty preset.
%Respectively, we denote by $\mathit{Max}(N)$ the set of nodes of the net that has an empty postset.

\vspace{-2mm}
\begin{definition}[Branching process]\emph{
\label{def:branching:process} A \emph{branching process} of a system
$S=(N,M_0)$ is a pair $\beta=(N^\prime,\nu)$, where $N^\prime=(B,E,G)$ is an
occurrence net and $\nu$ is a homomorphism from $N^\prime$ to $N$, such
that:
\begin{compactitem}
  \item the restriction of $\nu$ to $\mathit{Min}(N^\prime)$ is a bijection
between $\mathit{Min}(N^\prime)$ and $M_0$, and
  \item for all $e_1,e_2 \in E$ holds if $\pre{e_1}=\pre{e_2}$ and
$\nu(e_1)=\nu(e_2)$, then $e_1 = e_2$.
\end{compactitem}
}\end{definition}
\vspace{-2mm}

\noindent The system $S$ is referred to as the \emph{originative} system of a branching process. A branching process can be a \emph{prefix} of another branching process.
%Two branching processes $\beta_1 = (N_1,\nu_1)$ and $\beta_2 = (N_2,\nu_2)$ are \emph{isomorphic} if there is a bijective homomorphism from $N_1$ to $N_2$.
%Alternatively, two branching processes are isomorphic if they differ only in the names of nodes.

\vspace{-2mm}
\begin{definition}[Prefix]\emph{
\label{def:prefix} Let $\beta_1 = (N_1,\nu_1)$ and $\beta_2 = (N_2,\nu_2)$
be
two branching processes of a system $S=(N,M_0)$. $\beta_1$ is a
\emph{prefix}
of $\beta_2$ if $N_1$ is a subnet of $N_2$, such that:
if a condition belongs to $N_1$, then its input event in $N_2$ also
belongs to $N_1$,
if an event belongs to $N_1$, then its input and output conditions in
$N_2$ also belong to $N_1$, and $\nu_1$ is the restriction of $\nu_2$ to
nodes of $N_1$.}
\end{definition}
\vspace{-2mm}

\noindent
A maximal branching process of $S$ with respect to the prefix relation is
called \emph{unfolding} of the system.
%In~\cite{Engelfriet91}, it is shown that every system has a unique (up to isomorphism) unfolding.
Finally, we present a complete prefix unfolding.

%figure
%\figurename\ref{fig:unfolding} exemplifies the unfolding of a subnet $R1$ in \figurename~\ref{fig:wf:net} (\see dotted box). Observe that by $c_{x}, c^\prime_{x}, c^{\prime\prime}_{x}, \ldots \ $, we denote conditions that are the occurrences of place $p_x$ in the originative system. Similarly, by $e_{y}, e^\prime_{y}, e^{\prime\prime}_{y}, \ldots \ $, we denote events that are the occurrences of transition $t_y$ in the originative system. In the example, during the construction of a maximal branching process (an unfolding), after reaching conditions $c_w$ and $c_x$, it splits into two independent parallel copies of itself. Finally, we present the notion of a complete prefix unfolding.

\enlargethispage{\baselineskip}

\vspace{-2mm}
\begin{definition}[Complete prefix unfolding]\emph{
\label{def:configuration}
\newline
Let $\beta=(N^\prime,\nu)$, $N^\prime=(B,E,G)$, be a branching process of a
system $S=(N,M_0)$.
\begin{compactitem}
  \item A \emph{configuration} $C$ of $\beta$ is a set of events, $C
\subseteq E$, such that: (1) $e \in C$ implies that for all $e^\prime
\in E$, $e^\prime \leadsto e$ implies $e^\prime \in C$, \ie $C$
is causally closed, and (2) for all $e_1,e_2 \in C$ holds $\neg (e_1 \
\# \ e_2)$, \ie $C$ is conflict-free.
  \item A \emph{local} configuration of an event $e \in E$, denoted by
$\conf{e}$, is the set $\{e^\prime \in E \mid e^\prime \in E, e^\prime
\leadsto e \}$, \ie the set of events that precede $e$.
  \item A set of conditions of an occurrence net is a \emph{co-set} if its
elements are pairwise concurrent. A maximal co-set with respect to
inclusion is a \emph{cut}.
  \item For a finite configuration $C$ of $\beta$,
$\mathit{Cut}(C)=(\mathit{Min}(N^\prime) \cup \post{C}) \setminus
\pre{C}$ is a cut, whereas $\nu(\mathit{Cut}(C))$ is a reachable
marking of $S$, denoted by $\mathit{Mark}(C)$.
  \item $\beta$ is \emph{complete} if for each reachable marking $M$ of $S$
there exists a configuration $C$ in $\beta$, such that: (1)
$\mathit{Mark}(C) = M$, \ie $M$ is represented in $\beta$, and (2) for
each transition $t$ enabled at $M$ in $N$, there exists a configuration
$C \cup \{e\}$ in $\beta$, such that $e \notin C$ and $\nu(e) = t$.
  \item An \emph{adequate order} $\order$ is a strict well-founded partial
order on local configurations, such that $\conf{e} \subset
\conf{e^\prime}$ implies $\conf{e} \order \conf{e^\prime}$, where
$e,e^\prime \in E$.
  \item An event $e \in E$ is a \emph{cutoff} event induced by $\order$,
\ifaof there exists a \emph{corresponding} event $e^\prime \in E$, or
$corr(e)$, such that $\mathit{Mark}(\conf{e}) =
\mathit{Mark}(\conf{e^\prime})$ and $\conf{e^\prime} \order \conf{e}$.
  \item $\beta$ is the \emph{complete prefix unfolding} induced by $\order$,
\ifaof $\beta$ is the greatest prefix of the unfolding of $S$ that
contains no event after a cutoff event.
  %\item A \emph{complete prefix unfolding} is the greatest backward closed subnet of an unfolding that contains no events after cutoff events.
\end{compactitem}
}\end{definition}
\vspace{-2mm}

\noindent
The complete prefix unfolding is obtained by truncating the unfolding at events

\begin{wrapfigure}{r}{.5\textwidth}
\vspace{-4mm}
\begin{center}
  \includegraphics[scale = .48]{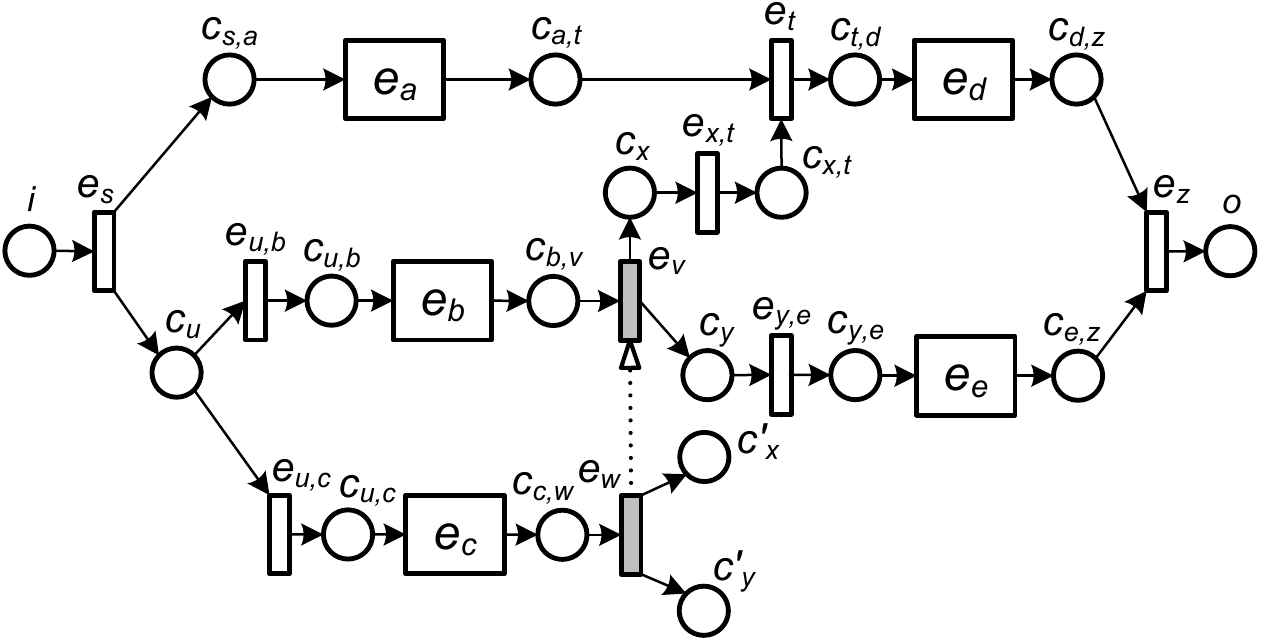}
\end{center}
\vspace{-7mm}
\caption{\small{A complete prefix unfolding of the system in
\figurename~\ref{fig:net}}} 	\label{fig:unfolding}
\vspace{-8mm}
\end{wrapfigure}

\noindent
where the information about reachable markings starts to be redundant. \figurename~\ref{fig:unfolding} shows a complete prefix unfolding of the system in \figurename~\ref{fig:net}.
In the prefix, event $e_w$ is a cutoff event, whereas event $e_v$ is its corresponding event; this relation is visualized by a dotted arrow. We write $c_{x}, c^\prime_{x}, c^{\prime\prime}_{x}, \ldots \ $ for conditions that are the occurrences of place $p_x$; correspondingly for events.
The size of the prefix depends on the ``quality'' of the adequate order used to perform the truncation.
It has been shown that the adequate order proposed in~\cite{EsparzaRV02} results in more compact prefixes as compared to the one in~\cite{McMillan95}.

%%%%%%%%%%%%%%%%%%%%%%%%%%%%%%%%%%%%%%%%%%%%%%%%%%%%%%%%%%%%%%%%%%%%%%%%%%%%%%%%%%%%%%%%%%

%% file: structuring.tex
%%%%%%%%%%%%%%%%%%%%%%%%%%%%%%%%%%%%%%%%%%%%%%%%%%%%%%%%%%%%%%%%%%%%%%%%%
\vspace{-3mm}
\section{Structuring}
\label{sec:structuring}
\vspace{-2mm}
%%%%%%%%%%%%%%%%%%%%%%%%%%%%%%%%%%%%%%%%%%%%%%%%%%%%%%%%%%%%%%%%%%%%%%%%%

\enlargethispage{\baselineskip}

% Intro
This section discusses the technique for structuring acyclic process models, 
presented in~\cite{PGD10}.
We elaborate further on the technique by proposing the notion of a proper 
complete
prefix unfolding for the first time; we will see that this prefix is
essential to achieve \emph{maximal} structuring.

% A process model
\begin{wrapfigure}{r}{.5\textwidth}
\vspace{-7mm}
\begin{center}
  \includegraphics[scale=.47, trim = 2mm 0 0 0]{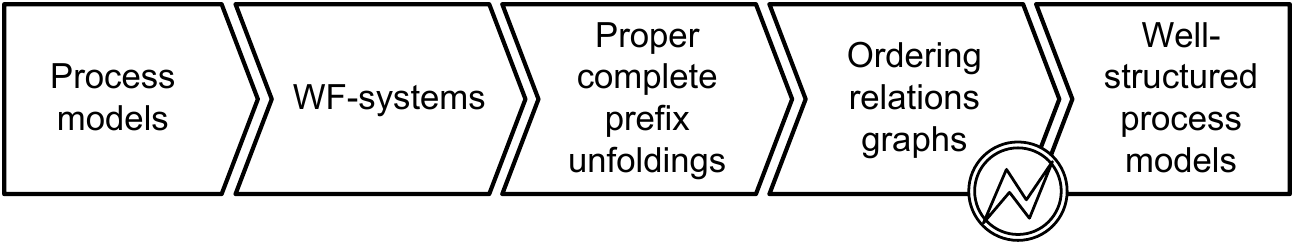}
\end{center}
\vspace{-6mm}
  \caption{\small{Structuring chain,~\seeCite{PGD10}}}
  \label{fig:bpm10}
\vspace{-6mm}
\end{wrapfigure}

\figurename~\ref{fig:bpm10} shows a chain of phases that collectively 
compose
the structuring technique. The process model is decomposed into a hierarchy
of process \emph{components}. Each component is a process model by itself 
and
either well-structured or unstructured. An unstructured process component 
can
in some cases be transformed into a well-structured one. For this purpose,
the component is translated into a workflow system for which the ordering
relations of its tasks are derived from its proper complete prefix
unfolding. If the ordering relations have certain properties, the
unstructured component can be replaced by a well-structured hierarchy of
smaller components that define the same ordering relations.
%The structuring in~\cite{PGD10} fails if a given process model has no equivalent well-structured representation. The model may, however, be partially structured into an equivalent maximally-structured version, as in \figurename~\ref{fig:and:prim}.
In the following, we present each phase of
the structuring in detail, whereas in the next section we extend
the technique to allow maximal structuring.

%%%%%%%%%%%%%%%%%%%%%%%%%%%%%%%%%%%%%%%%%%%%%%%%%%%%%%%%%%%%%%%%%%%%%%%%%%%%%%%%%%%%%%%%%%
\vspace{-4mm}
\subsubsection{From process models to unfoldings.}
%%%%%%%%%%%%%%%%%%%%%%%%%%%%%%%%%%%%%%%%%%%%%%%%%%%%%%%%%%%%%%%%%%%%%%%%%%%%%%%%%%%%%%%%%%

% Process model and RPST
\figurename~\ref{fig:structurable} shows a process model that will be used in
this section for explaining the structuring technique. We employ the Refined
Process Structure Tree (RPST)~\cite{PVV10,VVK09} to learn its structural
characteristics. The RPST is built from four kinds of \emph{process
components}: A \emph{trivial} (T) component consists of a single flow arc. A
\emph{polygon} (P) represents a sequence of components. A \emph{bond} (B)
stands for a set of components that share two common nodes -- an entry and
exit. Any other component is a \emph{rigid} (R). A component is
\emph{canonical}, \ifaof it does not overlap (on edges) with any other
component. The set of all canonical process components forms a hierarchy that
can be represented as a tree -- the RPST. The parent of a process component
is the smallest component that contains it. The root of the RPST captures the
whole process model, and a leaf of the RPST is a flow arc. The dotted boxes
in \figurename~\ref{fig:structurable} indicate the components and their
hierarchy, \eg $P1$ is a polygon which consists of trivial components
$(i,s)$, $(z,o)$, and rigid $R1$. Observe that we do not explicitly visualize
\emph{simple} components, \ie trivials and polygons composed of two trivials.

% A process model
\begin{wrapfigure}{r}{.5\textwidth}
\vspace{-10mm}
\begin{center}
  \includegraphics[scale=.6]{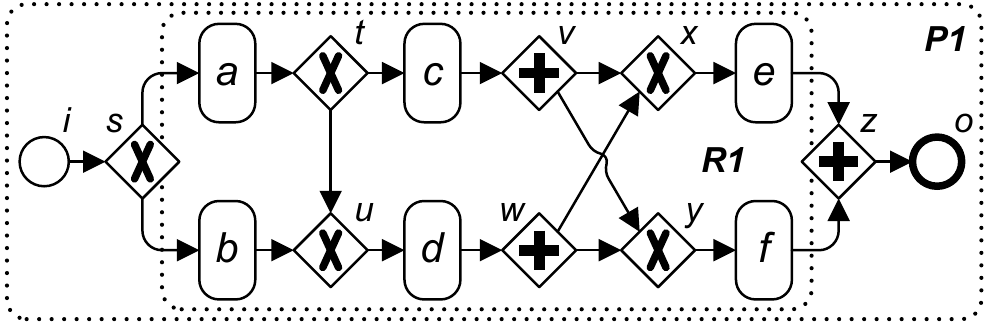}
\end{center}
\vspace{-6mm}
  \caption{\small{A process model}}
  \label{fig:structurable}
\vspace{-8mm}
\end{wrapfigure}

% RPST amd well-structuredness
Polygons and bonds correspond to sequences and well-structured components of
mutually-exclusive or concurrent threads. Therefore, a process model is
well-structured, \ifaof its RPST contains no rigid components. A process 
model can be structured by traversing its RPST bottom-up and
replacing each rigid component by its equivalent well-structured component.
The difficult step is to find this equivalent well-structured component.

The key idea of structuring is to \emph{refine} a rigid component $R$, \ie
a \emph{node} of the RPST, by a subtree of well-structured RPST nodes which 
define the same
behavioral relations between $R$'s children.
%; the subtree defines the structured representation of $R$.
The first step when structuring a rigid component is to compute the
ordering relations of its child nodes. We obtain these by constructing a
complete prefix unfolding of $R$'s corresponding WF-system. The complete
prefix unfolding captures information about all reachable markings of the
originative system, but has a simpler structure, \ie it is an
occurrence net (Def.~\ref{def:occ:net}).
%defining the ordering relations.
To capture all well-structuredness contained in $R$, the complete
prefix unfolding must have a specific shape called \emph{proper}.

\enlargethispage{\baselineskip}

\vspace{-1.5mm}
\begin{definition}[Proper complete prefix unfolding]\emph{
\label{def:pcpu} Let $\beta=(N^\prime,\nu)$, $N^\prime=(B,E,G)$, be a
branching process of an acyclic system $S=(N,M_0)$.
\begin{compactitem}
  \item A cutoff event $e \in E$ of $\beta$ induced by an adequate order
$\order$ is \emph{healthy}, \ifaof $\mathit{Cut}(\conf{e}) \setminus
\post{e} = \mathit{Cut}(\conf{\corr{e}}) \setminus \post{\corr{e}}$.
  \item $\beta$ is the \emph{proper complete prefix unfolding}, or the
\emph{proper prefix}, induced by an
adequate order $\order$, \ifaof $\beta$ is the greatest prefix of the
unfolding of $S$ that contains no event after a healthy cutoff event.
\end{compactitem}
}\end{definition}
\vspace{-1.5mm}

% Proper complete prefix unfolding
\begin{wrapfigure}{r}{.5\textwidth}
\vspace{-12mm}
\begin{center}
  \includegraphics[scale=.45, trim = 5mm 0 0 0]{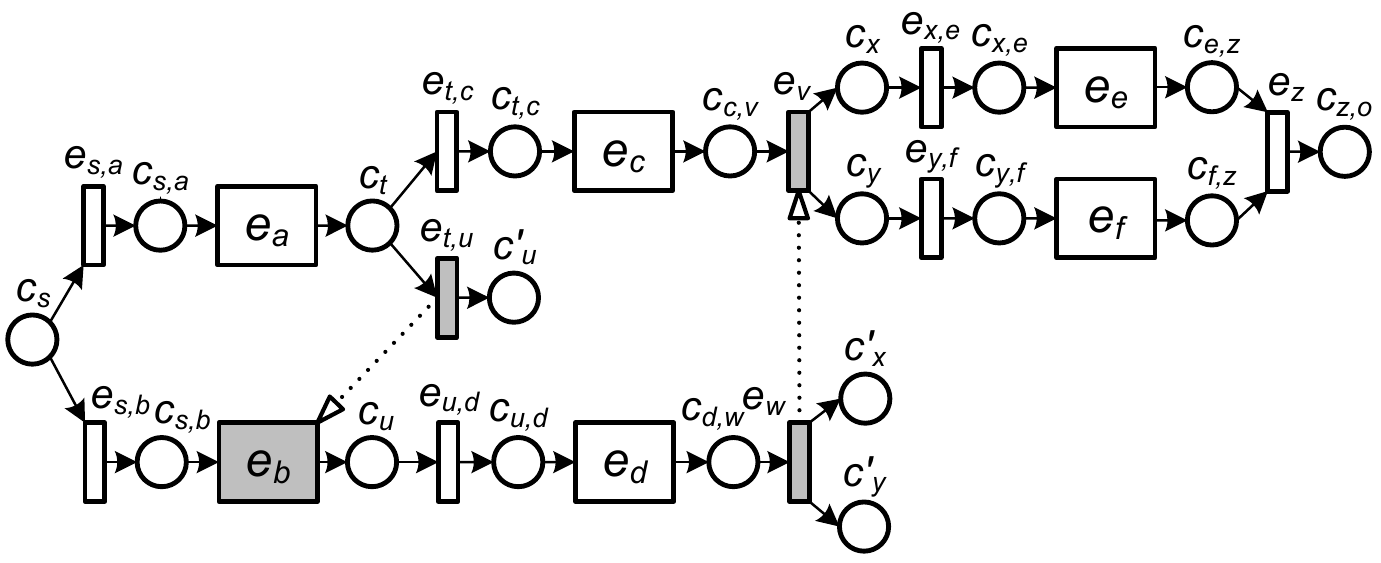}
\end{center}
\vspace{-6mm}
  \caption{\small{A proper complete prefix unfolding}}
  \label{fig:structurable:pcpu}
\vspace{-8mm}
\end{wrapfigure}

% Discussion on proper complete prefix unfolding
\noindent
\figurename~\ref{fig:structurable:pcpu} shows a proper prefix of the system 
that corresponds to the rigid component $R1$ in 
\figurename~\ref{fig:structurable}.
A proper prefix contains all information about well-structuredness,
\ie all paired gateways of splits and joins, in a rigid in the
following way. $\beta$ represents each \emph{xor} split as a condition
with multiple post-events; each \emph{xor} join is identified by the
post-conditions of a cutoff event $e$ and its corresponding event
$\corr{e}$, \eg $c_u$ and $c_u^\prime$ in
\figurename~\ref{fig:structurable:pcpu}. The notion of a cutoff event 
guarantees
that $\beta$ contains every \emph{xor} split and join. An important
observation here is that corresponding pairs of \emph{xor} splits and joins
are always contained in the same branch of $\beta$. An \emph{and} split 
manifests as an event with multiple post-conditions in $\beta$, whereas an
\emph{and} join is an event with multiple pre-conditions. The healthiness
requirement on cutoff events ensures that concurrency after an \emph{and}
split is kept encapsulated, \ie if several concurrent branches are 
introduced
in the unfolding they are not truncated until the point of their
synchronization, \ie the \emph{and} join. Such an intuition supports our 
goal
to derive a well-structured process model, as bonds of a process model that
define concurrency must be synchronized in the same branches of the model
where they originated.

A proper complete prefix unfolding of an acyclic system is clearly finite.
For structuring purposes, when computing a proper prefix, we use
an adequate order proposed in~\cite{EsparzaRV02}. This adequate order results 
in minimal complete prefix unfoldings for safe systems, if one only considers 
information about reachable markings induced by local configurations, which
is the case for healthy cutoff events. Thus, the adequate order 
from~\cite{EsparzaRV02} yields a minimal proper complete prefix unfolding of 
a
safe acyclic system, which applies to our case as sound free-choice nets are 
safe~\cite{A00}.

%%%%%%%%%%%%%%%%%%%%%%%%%%%%%%%%%%%%%%%%%%%%%%%%%%%%%%%%%%%%%%%%%%%%%%%%%%%%%%%%%%%%%%%%%%
\vspace{-4mm}
\subsubsection{From unfoldings to graphs.}
%%%%%%%%%%%%%%%%%%%%%%%%%%%%%%%%%%%%%%%%%%%%%%%%%%%%%%%%%%%%%%%%%%%%%%%%%%%%%%%%%%%%%%%%%%

\enlargethispage{\baselineskip}

The proper complete prefix unfolding of a process component $R$ contains all
ordering relations of all children of $R$ in the RPST. For restructuring, 
$R$
(an RPST node) is to be refined into a subtree along these ordering
relations. The refinement requires this information to be preserved
in a hierarchically decomposable form: an ordering relations \emph{graph}.

% Ordering relations graph
\vspace{-1.5mm}
\begin{definition}[Ordering relations graph]\emph{
\label{def:org}
\newline
Let $\beta=(N^\prime,\nu)$, $N^\prime=(B,E,G)$, be a proper complete prefix
unfolding of a sound acyclic free-choice WF-system $S=(N,M_i)$,
$N=(P,T,F,\mathcal{T},\lambda)$.
\begin{compactitem}
  \item % proper causal
Two nodes $x$ and $y$ of $N^\prime$ are in \emph{proper causal} relation,
denoted by $x \rightarrowtail y$, \ifaof $(x,y) \in G^+$ or there exists a
sequence $(e_1,\ldots,e_n)$ of proper cutoff events of $\beta$, $e_i \in E$,
$1 \leq i \leq n$, $n \in \mathbb{N}$, such that $(x,e_1) \in G^\ast$,
$(\corr{e_n},y) \in G^+$, and $(\corr{e_i},e_{i+1}) \in G^\ast$ for $1 \leq
i
< n$. We denote by $\leftarrowtail$ the inverse of $\rightarrowtail$.
  \item % proper relations
Let $\mathcal{R}=\{\leadsto, \leftrsquigarrow, \#, ||\}$ be the ordering
relations of $N^\prime$. The \emph{proper conflict} relation of $N^\prime$
is
$\mathord{\boxplus} = \mathord{\#} \setminus (\rightarrowtail \cup 
\leftarrowtail)$. The set
$\mathcal{R}^\prime=\{\rightarrowtail, \leftarrowtail, \boxplus, || \}$
forms
the \emph{proper ordering relations} of $N^\prime$.
  \item % observable proper relations
We refer to $\mathcal{R}$ as \emph{observable} (proper) ordering relations, 
\ifaof
the relations in $\mathcal{R}$ only contain pairs of events that correspond
to observable transitions of $N$.
  \item % ordering relations graph
Let $\mathcal{R}=\{\rightarrowtail, \leftarrowtail, \boxplus, || \}$ be the
observable proper ordering relations of $N^\prime$. An \emph{ordering
relations graph} $\mathcal{G}=(V,A,\mathcal{B},\sigma)$ of $N^\prime$ has
vertices $V \subseteq E$ defined by events of $\beta$ that correspond to
observable transitions of $N$, arcs $A = \ \rightarrowtail \cup \ \boxplus$,
and a labeling function $\sigma : V \rightarrow \mathcal{B}, \mathcal{B} =
\mathcal{T} \setminus \{ \tau \}$ with $\sigma(v)=\lambda(\nu(v))$, $v \in
V$.
\end{compactitem}
}\end{definition} \vspace{-1.5mm}

\begin{figure}[h]
\vspace{-9mm}
\begin{center}
  \subfigure[]{\label{fig:structurable:org}\includegraphics[scale = .65,
trim
= 0 -4mm 0 0]{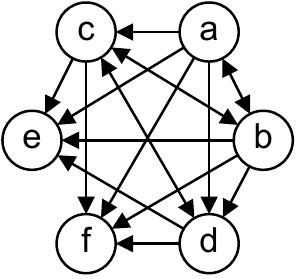}}
  \hspace{3mm}
  \subfigure[]{\label{fig:structurable:mdt}\includegraphics[scale =
.62]{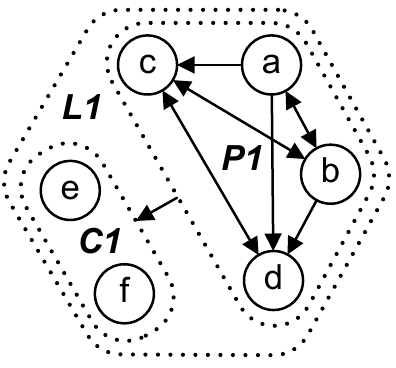}}
  \hspace{3mm}
  \subfigure[]{\label{fig:and:prim:org}\includegraphics[scale = .7, trim = 0
-4mm 0 0]{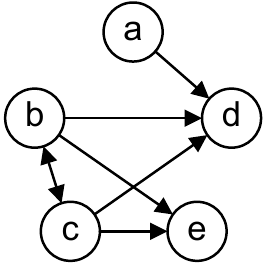}}
  \hspace{3mm}
  \subfigure[]{\label{fig:and:prim:mdt}\includegraphics[scale = .7, trim = 0
-1mm 0 0]{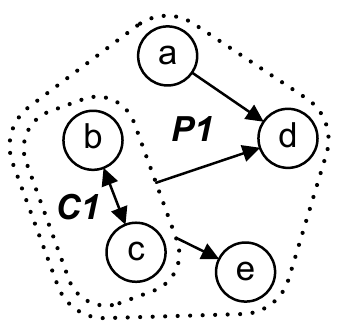}}
\end{center}
\vspace{-7mm}
\caption{\small{(a),(c) Ordering relations graphs, and (b),(d)
the modular decomposition trees}} \label{fig:mdts}
\vspace{-7mm}
\end{figure}

\noindent An ordering relations graph of a process component $R$ captures
minimal and complete information about the ordering relations of events that
correspond to observable transitions of a system.
\figurename~\ref{fig:structurable:org}
visualizes the ordering relations graph of the proper complete prefix
unfolding in \figurename~\ref{fig:structurable:pcpu}. The proper causal
relation $\rightarrowtail$ updates the causality relation of the prefix
$\beta$ to overcome the effect of unfolding truncation, \eg $a
\rightarrowtail d$, $b \rightarrowtail f$, $d \rightarrowtail e$, etc. 
%Due to a design decision, arcs of an ordering relations graph encode proper causal and proper conflict relations.
\figurename~\ref{fig:structurable:org} denotes that $a$ and $c$ are
in proper causal relation, $a$ and $b$ are in proper conflict, whereas $e$
and $f$ are concur-

\noindent
rent. Figures~\ref{fig:and:prim:org} and~\ref{fig:and:prim:mdt} show the graph and its MDT of the
model in \figurename~\ref{fig:and:prim:unstr}.

%%%%%%%%%%%%%%%%%%%%%%%%%%%%%%%%%%%%%%%%%%%%%%%%%%%%%%%%%%%%%%%%%%%%%%%%%%%%%%%%%%%%%%%%%%
\vspace{-4.5mm}
\subsubsection{From graphs to process models.}
%%%%%%%%%%%%%%%%%%%%%%%%%%%%%%%%%%%%%%%%%%%%%%%%%%%%%%%%%%%%%%%%%%%%%%%%%%%%%%%%%%%%%%%%%%

\enlargethispage{\baselineskip}

The ordering relations graph not only encodes the ordering relations, it also
inherits \emph{all} information about well-structured\-ness from the proper
prefix, \ie pairing of gateways
is preserved. The structuring technique in~\cite{PGD10}
proceeds by parsing the graph into a hierarchy of subgraphs that encode
ordering relations of well-structured components. The thereby
discovered hierarchy of subgraphs is then used to refine a rigid component
into a subtree. As shown in~\cite{PGD10},
each subgraph corresponds to the notion of a \emph{module} of the modular
decomposition of a directed graph~\cite{McConnellM05} -- thus discovering
well-structuredness in the relations of an unstructured process component.

% Modules
Let $\mathcal{G}=(V,A,\mathcal{B},\sigma)$ be an ordering relations graph. A
\emph{module} $M \subseteq V$ in $\mathcal{G}$ is a non-empty subset of
vertices of $\mathcal{G}$ that are in uniform relation with vertices $V
\setminus M$, \ie if $v \in V  \setminus M$, then $v$ has directed edges to
all members of $M$ or to none of them, and all members of $M$ have directed
edges to $v$ or none of them do. However, $v_1,v_2 \in V \setminus M$, $v_1
\neq v_2$ can have different relations to members of $M$. Moreover, the
members of $M$ and $V \setminus M$ can have arbitrary relations to each
other~\cite{McConnellM05}. For example, $\{e,f\}$ is a module
in \figurename~\ref{fig:structurable:org}.
%Let $\mathcal{G}$ be an ordering relations graph.
Two modules $M_1$ and $M_2$
of $\mathcal{G}$ \emph{overlap}, \ifaof they intersect and neither is a
subset of the other, \ie $M_1 \setminus M_2$, $M_1 \cap M_2$ , and $M_2
\setminus M_1$ are all non-empty. $M_1$ is \emph{strong}, \ifaof there 
exists
no module $M_2$ of $\mathcal{G}$, such that $M_1$ and $M_2$ overlap. The
\emph{Modular Decomposition Tree} (MDT) of $\mathcal{G}$ is a set of all
strong modules of $\mathcal{G}$. The modular decomposition substitutes
each strong module of $\mathcal{G}$ by a new vertex and proceeds 
recursively.
The result is the MDT which is a canonical rooted tree and unique.
%The MDT of a directed graph can be computed in linear time~\cite{McConnellM05}.

Now, a rigid process component $R$ of an RPST can be restructured by 
refining
$R$ in the RPST to a subtree $T_R$. The root of $T_R$ is child of $R$'s
parent, each child of $R$ is attached to a leaf of $T_R$, the nodes of $T_R$
are defined by the modules of the MDT of $R$'s ordering relations graph. The
type of a node of $T_R$ is determined by the characteristics of its defining
MDT module, as follows.

We refer to singletons of $V$ as the \emph{trivial} modules of 
$\mathcal{G}$.
Let $M$ be a non-trivial module.
$M$ is \emph{complete} ($C$), \ifaof the subgraph of $\mathcal{G}$ induced 
by
vertices in $M$ is either complete or edgeless. If the subgraph is complete,
then we refer to $M$ as $xor$ complete. If the subgraph is edgeless, then we
refer to $M$ as $and$ complete. $M$ is \emph{linear} ($L$), \ifaof there
exists a linear order $(x_1, \ldots, x_{|M|})$ of elements of $M$, such that
there is a directed edge from $x_i$ to $x_j$ in $\mathcal{G}$, \ifaof $i<j$.
Finally, if $M$ is neither complete, nor linear, then $M$ is 
\emph{primitive}
($P$); a primitive module is \emph{concurrent} \ifaof it contains a pair of
vertices that are not connected by an edge.
\figurename~\ref{fig:structurable:mdt} shows the MDT of the graph
in~\figurename~\ref{fig:structurable:org}. Besides the trivial modules, the
MDT contains linear $L1$, $and$ complete $C1$, and primitive $P1$. Module
$L1$ is the root module, whereas trivial modules are leafs of the MDT.

An acyclic process model has an equivalent well-structured model,
if its ordering relations graph contains no concurrent primitive module.
According to~\cite{PGD10}, behavior captured by other module classes can be
expressed by well-structured process components. A trivial module 
corresponds
to a task. A linear module corresponds to a polygon component. An $and$
($xor$) complete module corresponds to a bond with $and$ ($xor$)
gateways as entry and exit nodes.
%A primitive module without concurrency relations corresponds to some sequential system that can be structured by employing techniques from compiler theory.
A primitive module without concurrency can be restructured using standard 
compiler techniques~\cite{O82}.

Given all of the above, \algorithmname~\ref{algo:acyclic:structuring}
summarizes the structuring technique.

\vspace{-7mm}
\begin{algorithm}
\label{algo:acyclic:structuring} \caption{Structuring Acyclic Process Model
(Component)} \KwIn{An acyclic process model (component) $W$} \KwOut{A
well-structured process model that is equivalent to $W$} \BlankLine
Construct WF-net $N$ that corresponds to $W$\\
Construct proper complete prefix unfolding $\beta$ of $(N,M_i)$\\
Construct ordering relations graph $\mathcal{G}$ of $\beta$\\
Compute $\mathcal{M}$ -- the MDT of $\mathcal{G}$\\
\tcp{Construct process model $P$ by traversing $\mathcal{M}$ in postorder}
\ForEach{node $m$ of $\mathcal{M}$} {
  \lIf{$m$ is trivial}{Construct a task}\\
  \lIf{$m$ is $and$ complete}{Construct an $and$ bond component}\\
  \lIf{$m$ is $xor$ complete}{Construct a $xor$ bond component}\\
  \lIf{$m$ is linear}{Construct a trivial or polygon component}\\
  \If{$m$ is primitive without concurrency}
  {
    Construct a well-structured process component using compiler techniques
  }
  \lElse{FAIL}
} \Return $P$
\end{algorithm}
\vspace{-7mm}

\algorithmname~\ref{algo:acyclic:structuring} traverses the
MDT of an ordering relations graph of a rigid process component and 
constructs
for each encountered module a process component from components that
correspond to its child modules. The resulting hierarchy of components is 
the
subtree that refines the rigid component.

\begin{figure}[h]
\vspace{-8mm}
\begin{center}
  \subfigure[]{\label{fig:structurable:w:r}\includegraphics[scale = .6, trim
  = 0 -7mm 0 0]{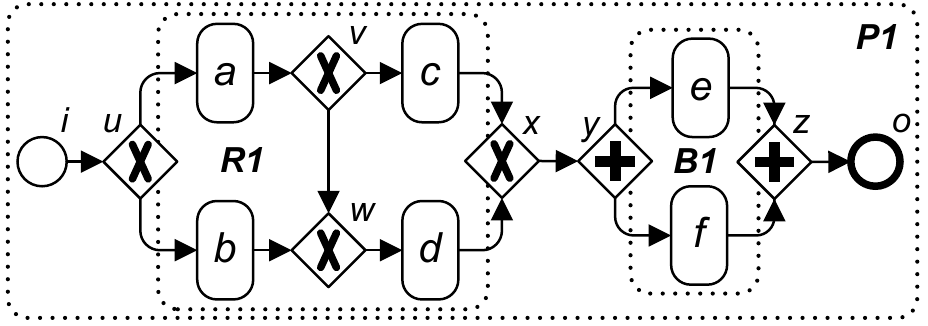}}
  \hspace{3mm}
  \subfigure[]{\label{fig:structurable:str}\includegraphics[scale =
.6]{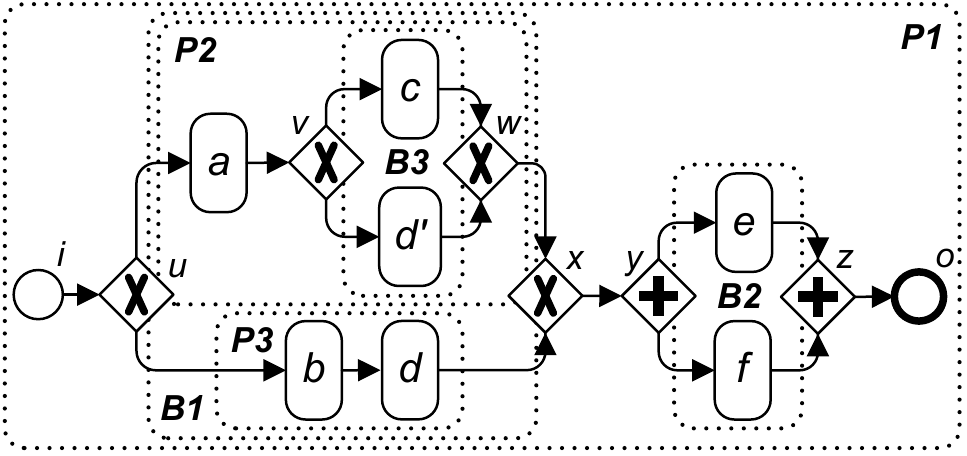}}
\end{center}
\vspace{-7mm} \caption{\small{(a),(b) Process models that are equivalent
with
the process model in \figurename~\ref{fig:structurable}}}
\label{fig:structurable:result} \vspace{-7mm}
\end{figure}

\figurename~\ref{fig:structurable:result} shows two process models that are
equivalent with the process model in \figurename~\ref{fig:structurable}. The
model in \figurename~\ref{fig:structurable:w:r} is obtained by constructing
process components that correspond to modules of the MDT in
\figurename~\ref{fig:structurable:mdt}. Here, polygon $P1$ corresponds to
linear $L1$, bond $B1$ to $and$ complete $C1$, and rigid $R1$ to primitive
$P1$. The model in \figurename~\ref{fig:structurable:str} is obtained from
\figurename~\ref{fig:structurable:w:r} by structuring rigid $R1$. The
structuring can be achieved by employing {\sffamily ID-0} transformation 
rule
from~\cite{O82}.

%% file: maximal.tex
\vspace{-3mm}
\section{Maximal Structuring}
\label{sec:maximal:structuring}
\vspace{-2mm}
%%%%%%%%%%%%%%%%%%%%%%%%%%%%%%%%%%%%%%%%%%%%%%%%%%%%%%%%%%%%%%%%%%%%%%%%%%%%%%%%%%%%%%%%%%

\enlargethispage{\baselineskip}

% Intro: Part I
Recall from Sect.~\ref{sec:introduction} that a process model is maximally
structured iff every equivalent model has the same number of process
components defined by pairs of splits and joins as the model itself. In the
light of Sect.~\ref{sec:structuring}, the open problem is to obtain a
maximally structured process component $R$. $R$ has this property iff (1) all
primitive modules in the MDT of $R$'s ordering relations graph are
concurrent, and (2) there exists a bijection between non-singleton modules of
the MDT and non-trivial components of the RPST which assigns to each
primitive module a rigid component, to each complete a bond, and to each
linear a polygon. The maximal structuredness of $R$ follows from the
maximality of the modular decomposition: the ordering relations graph of $R$
inherits all information about well-structuredness from the proper complete
prefix of $R$, and the MDT maximizes modules with a well-structured
representation because of the decomposition into strong modules. If a
primitive module $M$ with concurrency has well-structured child modules, then
these modules are maximal again within $M$. Only the relations within $M$
have no structured representation as a process model, where $M$ is minimized
by maximizing structuredness around and inside $M$. This yields a technique
for maximal structuring: one must be able to synthesize a process component
that exhibits the ordering relations described in $M$. Such a technique would
allow to define unstructured process model topologies when mapping
hierarchies of modules onto hierarchies of process components in
\algorithmname~\ref{algo:acyclic:structuring}, \eg primitive modules in
\figurename~\ref{fig:structurable:mdt} and \figurename~\ref{fig:and:prim:mdt}
onto rigid components in \figurename~\ref{fig:structurable:w:r} and
\figurename~\ref{fig:and:prim:str}. The resulting process model would be
maximally structured.

% Intro: Part I
%In order to achieve the maximal structuring of a process model one must
%possess a technique that is a ``reverse'' of the structuring technique
%proposed in \sectionname~\ref{sec:structuring}. Specifically, one must be
%able to synthesize a process model that exhibits ordering relations described
%in a given ordering relations graph. Such a technique would allow definitions
%of unstructured process model topologies when mapping hierarchies of modules
%onto hierarchies of process components, \eg primitive modules in
%\figurename~\ref{fig:structurable:mdt} and \figurename~\ref{fig:and:prim:mdt}
%onto rigid components in \figurename~\ref{fig:structurable:w:r} and
%\figurename~\ref{fig:and:prim:str}. Such a mapping results in a
%\emph{maximally-structured} process model, for the following reasons: (i) The
%RPST defines the most fine-grained decomposition of a process model into
%process components. (ii) The ordering relations graph of an unstructured
%component $R$ contains all information about well-structuredness in $R$,
%because it is based on the proper complete prefix of $R$. (iii) The MDT
%maximizes modules with a well-structured representation because of the
%decomposition into strong modules. (iv) If a primitive module $M$ with
%concurrency has well-structured child modules, then these modules are maximal
%again within $M$. Only the relations within $M$ have no structured
%representation as a process model, where $M$ is minimized by maximizing
%structuredness around and inside $M$.

\begin{figure}[h]
\vspace{-4mm}
\begin{center}
  \includegraphics[width=\linewidth]{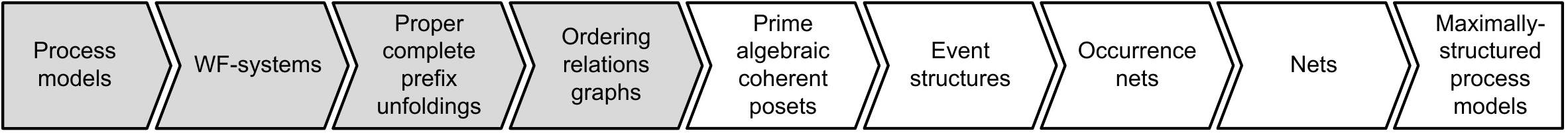}
\end{center}
\vspace{-6mm}
  \caption{\small{An extension of the structuring chain
of~\figurename~\ref{fig:bpm10}}}
  \label{fig:bpm11}
\vspace{-6mm}
\end{figure}

% Intro: Part II
In this section, we propose a solution to the synthesis problem, \ie given an
ordering relations graph (a module of an MDT) we synthesize a process model
(a component of the RPST) that realizes the relations described in
the graph. The solution consists of several phases that employ the results
on translations between the languages of domain and net
theory~\cite{NielsenPW81}, and on folding prefixes of
systems~\cite{FahlandThesis}. \figurename~\ref{fig:bpm11} shows an extension
of the structuring approach which was proposed
in~\figurename~\ref{fig:bpm10}. Next, we discuss each phase of
the extension in detail.

\vspace{-4mm}
\subsubsection{From graphs to partial orders.}
%%%%%%%%%%%%%%%%%%%%%%%%%%%%%%%%%%%%%%%%%%%%%%%%%%%%%%%%%%%%%%%%%%%%%%%%%%%%%%%%%%%%%%%%%%

\enlargethispage{\baselineskip}

This section describes a translation from an ordering relations graph to a
partial order of information. The partial order is an alternative
formalization of the meaning of the behavior captured in the graph.

% An ordering relations graph
\begin{wrapfigure}{r}{.215\textwidth}
\vspace{-10mm}
\begin{center}
  \includegraphics[scale=.7]{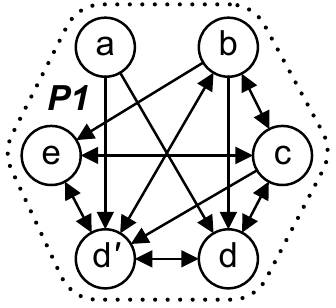}
\end{center}
\vspace{-7mm}
  \caption{\small{An order\-ing relations graph}}
  \label{fig:running:org}
\vspace{-8mm}
\end{wrapfigure}

The ordering relations graph in \figurename~\ref{fig:running:org} is the
running example of Sect.~\ref{sec:maximal:structuring}. The graph
is a primitive module with all types of relations;
$d$ and $d^\prime$ are events with the same label.

% POSETs: Part I
First, we give some definitions from the theory of partially ordered sets
(posets)~\cite{NielsenPW81}.
Let $(D,\sqsubseteq)$ be a poset.
% upper/lower bounds
For a subset $X$ of $D$, an element $y \in D$ is an upper (lower) bound of
$X$, \ifaof $x \sqsubseteq y$ ($x \sqsupseteq y$), for each element $x \in
X$.
% greatest/least element
An element $y \in D$ is a greatest (least) element, \ifaof for each element
$x \in D$ holds $x \sqsubseteq y$ ($x \sqsupseteq y$).
% maximal/minimal elements
An element $y \in D$ is a maximal (minimal) element, \ifaof there exist no
element $x \in D$, such that $y \sqsubset x$ ($x \sqsubset y$); $D_{max}$ and
$D_{min}$ denote the sets of maximal and minimal elements of $D$.
% POSETs: Part II
% compatible
Two elements $x$ and $y$ in $D$ are \emph{consistent}, written $x \uparrow
y$, \ifaof they have an upper bound, \ie $x \uparrow y \Leftrightarrow
\exists \ z \in D : x \sqsubseteq z \land y \sqsubseteq z$; otherwise they
are \emph{inconsistent}.
% pairwise consistent
A subset $X$ of $D$ is \emph{pairwise consistent}, written $X^\Uparrow$,
\ifaof every two elements in $X$ are consistent in $D$, \ie $X^\Uparrow
\Leftrightarrow \forall x,y \in X : x \uparrow y$.
% coherent
The poset $(D,\sqsubseteq)$ is \emph{coherent}, \ifaof each pairwise
consistent subset $X$ of $D$ has a least upper bound (lub) $\sqcup X$.
% complete prime
An element $x \in D$ is a \emph{complete prime}, \ifaof for each subset $X$
of $D$, which has a lub $\sqcup X$, holds that $x \sqsubseteq \sqcup X
\Rightarrow \exists \ y \in X : x \sqsubseteq y$. Let 
$P=(D,\sqsubseteq)$ be
a poset. We write $\mathfrak{P}_P$ for the set of complete primes of $P$.
% prime algebraic
The poset $P=(D,\sqsubseteq)$ is \emph{prime algebraic}, \ifaof
$\mathfrak{P}_P$ is denumerable and every element in $D$ is the lub of the
complete primes it dominates, \ie $\forall \ x \in D : x = \sqcup \{y \ | \ y
\in \mathfrak{P}_P \land y \sqsubseteq x \}$. A set $S$ is
\emph{denumerable}, \ifaof it is empty or there exists an enumeration of $S$
that is a surjective mapping from the set of positive integers onto $S$.

% Posets
\begin{wrapfigure}{r}{.41\textwidth}
\vspace{-12mm}
\begin{center}
  \subfigure[]{\label{fig:running:poset}\includegraphics[scale = .72, trim = 0 -8.6mm 0 0]{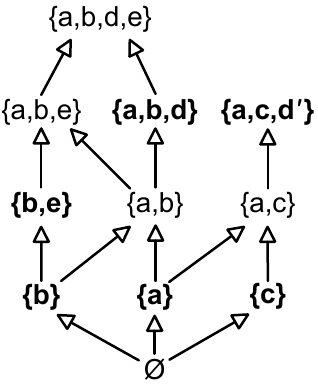}}
  \subfigure[]{\label{fig:running:poset:aug}\includegraphics[scale = .72]{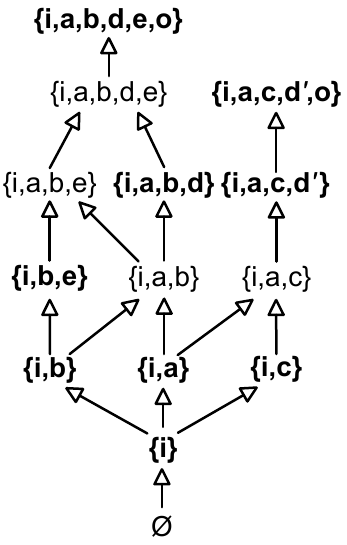}}
\end{center}
\vspace{-6mm}
  \caption{\small{(a) Poset, and (b) augmen\-ted poset obtained from \figurename~\ref{fig:running:org}}}
  \label{fig:running:posets}
\vspace{-7mm}
\end{wrapfigure}

% Towards L[G]
The behavior captured in an ordering relations graph can be given as a partial order of information points.
Similar to~\cite{NielsenPW81}, the
information points are chosen to be left-closed and conflict-free subsets of
vertices of the graph. Each such set captures the history of
events of some run of a system. Let $\mathcal{G}=(V,A,\mathcal{B},\sigma)$ be
a graph and let $W$ be a subset of $V$. $W$ is
\emph{conflict-free}, \ifaof $\forall \ v_1,v_2 \in W : (v_1,v_2) \nin A \lor
(v_2,v_1) \nin A$. $W$ is \emph{left-closed}, \ifaof $\forall \ v_1 \in W \
\forall \ v_2 \in V : (v_2,v_1) \in A \land (v_1,v_2) \notin A \Rightarrow
v_2 \in W$. We define $\mathcal{L}[\mathcal{G}]$ as the partial order of
left-closed and conflict-free subsets of $V$, ordered by inclusion.
\figurename~\ref{fig:running:poset} shows $\mathcal{L}$ of the graph in
\figurename~\ref{fig:running:org}. Thm.~\ref{thm:org:2:poset}, inspired by Thm.~8 in~\cite{NielsenPW81}, characterizes the posets
$\mathcal{L}[\mathcal{G}]$.

\enlargethispage{\baselineskip}

\vspace{-2mm}
\begin{theorem}
\label{thm:org:2:poset} Let $\mathcal{G}=(V,A,\mathcal{B},\sigma)$ be an
ordering relations graph. Then, $\mathcal{L}[\mathcal{G}]=(H,\subseteq)$ is a prime
algebraic coherent partial order. Its complete primes are those elements of
the form $[v]=\{v^\prime \in V \ | \ (v^\prime,v) \in B^\ast \}$, where $B=\{
a \in A \ | \ a^{-1} \notin A \}$.
\end{theorem}
\vspace{-4mm}
\begin{proof}
% coherent
Let $X \subseteq H$ be pairwise consistent. Then,
$\cup X$ is conflict-free. $\sqcup X = \cup X$ and, hence,
$\mathcal{L}[\mathcal{G}]$ is coherent.
% complete primes
Each $[v]$, $v \in V$, is clearly left-closed and conflict-free.
Let $X \subseteq H$ have lub $\sqcup X$.
$X$ is pairwise consistent and $\sqcup X = \cup X$.
Each $[v]$ is a complete prime.
If $[v] \subseteq \cup X$, then $v \in \cup X$ and for some $x \in X$ holds $v \in x$ and, thus, $[v] \subseteq x$.
% prime algebraic
It holds for each $X \in H$ that $X= \cup \{[v] \ | \ v \in \cup X \}$.
Thus, each element of $\mathcal{L}[\mathcal{G}]$ is a lub of the complete primes below it.
\qed
\end{proof}
\vspace{-2mm}

\noindent Given an ordering relations graph, one can construct
$\mathcal{L}[\mathcal{G}]=(H,\subseteq)$ iteratively. Let $h_1$ and $h_2$ be
subsets of $V$, such that $h_2 \setminus h_1 = \{v\}$. Then $h_1,h_2 \in H$,
\ifaof $h_1=\emptyset$ or $\exists \ a \in h_1 : (v,a) \notin A$, and
$\forall \ b \in h_1 : (b,v) \notin A \lor (v,b) \notin A$, and $\forall \ c
\in V \setminus h_1, (c,v) \in A, (v,c) \notin A \ \exists \ d \in h_1 :
(c,d),(d,c) \in A$.
%\algorithmname~\ref{algo:org:2:poset} formalizes the construction.

%\todo{maybe Alg. can be removed}
% Algorithm
%\vspace{-6mm}
%\begin{algorithm}
%\label{algo:org:2:poset}
%\caption{From graphs to partial orders}
%\KwIn{An ordering relations graph $\mathcal{G}=(V,A,\mathcal{B},\sigma)$}
%\KwOut{A prime algebraic coherent partial order $\mathcal{L}[\mathcal{G}]=(H,\subseteq)$}
%\BlankLine
%$H = \{ \emptyset \}$ \tcp{Initialize $H$}
%$Q.push(\emptyset)$ \tcp{Initialize queue}
%\While{$\neg \ Q.empty()$}{
%$d_1 = Q.pop()$\\
%  \For{$v \in V \setminus d_1$}{
%    $d_2 = d_1 \cup \{v\}$\\
%    \If{$(d_1=\emptyset \lor \exists a \in d_1 : (v,a)\notin A) \land (\forall b \in d_1 : (b,v)\notin A \lor (v,b)\notin A) \land
%    				(\forall c \in V \setminus d_1, (c,v) \in A, (v,c) \notin A \ \exists d \in d_1 : (c,d),(d,c) \in A)$}{
%      $H = H \cup \{ d_2 \}$\\
%      $Q.push(d_2)$
%    }
%  }
%}
%\Return $(H,\subseteq)$
%\end{algorithm}
%\vspace{-6mm}

% Augmented L[G]
Let $\mathcal{L}[\mathcal{G}]=(H,\subseteq)$ be a partial order of an
ordering relations graph. We augment $\mathcal{L}[\mathcal{G}]$ with two
fresh events $i,o \notin V$. These events are designed to ensure the
existence of a single source and single sink. An \emph{augmented} partial
order of

\noindent%%
$\mathcal{G}$ is $\mathcal{L}^\ast[\mathcal{G}] = (H^\ast,\subseteq)$, where
$H^\ast = \emptyset \cup \{h \cup \{i\} \ | \ h \in H \} \cup \{ h \cup
\{i,o\} \ | \ h \in H_{max} \}$. \figurename~\ref{fig:running:poset:aug}
shows $\mathcal{L}^\ast$ of the graph in \figurename~\ref{fig:running:org}.
After adding the minimal and maximal elements, the topology of posets stays
unchanged, so $\mathcal{L}^\ast$ is a prime algebraic coherent poset.

%%%%%%%%%%%%%%%%%%%%%%%%%%%%%%%%%%%%%%%%%%%%%%%%%%%%%%%%%%%%%%%%%%%%%%%%%%%%%%%%%%%%%%%%%%
\vspace{-4mm}
\subsubsection{From partial orders to event structures.}
%%%%%%%%%%%%%%%%%%%%%%%%%%%%%%%%%%%%%%%%%%%%%%%%%%%%%%%%%%%%%%%%%%%%%%%%%%%%%%%%%%%%%%%%%%

The next transformation step deals with translating partial orders to event
structures. Event structures are intermediate concepts between partial orders
and occurrence nets. The use of this intermediate concept was extensively
studied in~\cite{NielsenPW81}.

% Event structure
\vspace{-2mm}
\begin{definition}[Labeled event structure]\emph{
\label{def:event:structure}
\newline
An \emph{event structure} is a triple $\mathcal{E}=(E,\leq,\oplus)$, where
$E$ is a set of events, $\leq$ is a partial order over $E$ called the
causality relation, and $\oplus$ is a symmetric and irreflexive relation in
$E$, called the conflict relation that satisfies the principle of
\emph{conflict heredity}, \ie $\forall e_1,e_2,e_3 \in E : e_1 \geq e_2
\oplus e_3 \Rightarrow e_1 \oplus e_3$. A \emph{labeled} event structure
$\mathcal{E}=(E,\leq,\oplus,\mathcal{C},\kappa)$ additionally has a set
$\mathcal{C}$ of \emph{labels}, $\tau \in \mathcal{C}$, and $\kappa : E
\rightarrow \mathcal{C}$ assigns to each event a label.}
\end{definition}
\vspace{-2mm}

\enlargethispage{\baselineskip}

% An event structure [OLD]
%\begin{wrapfigure}{r}{.25\textwidth}
%\vspace{-1mm}
%\begin{center}
  %\includegraphics[scale=.75]{eventstr}
%\end{center}
%\vspace{-5mm}
  %\caption{\small{An event structure of \figurename~\ref{fig:running:poset}}}
  %\label{fig:running:eventstr}
%\vspace{-6mm}
%\end{wrapfigure}

\noindent%%%
An ordering relations graph $\mathcal{G}$ differs from an event structure
$\mathcal{E}$ in that $\mathcal{G}$ allows violations of
conflict heredity. These violations, however, are not harmful; they express
equivalent runs of a system. These equivalent run are visible in posets and
become explicit in event structures. A formal procedure for obtaining an
event structure from a graph can be intuitively understood as unfolding of
the graph. Next, we define a construction of an event structure from a poset.
The definition is an extension of Def.~18 in~\cite{NielsenPW81}; it
incorporates propagation of labels of an originative ordering relations graph
to the corresponding event structure.

\vspace{-2mm}
\begin{definition}[Event structure of partial order]\emph{
\label{def:poset:2:es}
\newline
Let $\mathcal{G}=(V,A,\mathcal{B},\sigma)$ be an ordering relations graph and
let $P=(H,\subseteq)$ be an (augmented) prime algebraic coherent partial
order of $\mathcal{G}$. Then, $\mathcal{P}[P]$ is defined as the labeled
event structure $(E,\leq,\oplus,\mathcal{C},\kappa)$, where $E =
\mathfrak{P}_P$, $\leq$ is $\subseteq$ restricted to $\mathfrak{P}_P$, for
all $e_1,e_2 \in \mathfrak{P}_P : e_1 \oplus e_2$, \ifaof $e_1$ and $e_2$ are
inconsistent in $P$, and $\mathcal{C} = \mathcal{B} \cup \{\tau\}$. Let $e
\in E$, and define $\hat{e}$ as $\hat{e} \in e \setminus \bigcup_{a \subset
e, a \in H} a$. Then, $\kappa(e)=\sigma(\hat{e})$, if $\hat{e} \in V$;
otherwise $\kappa(e)=\tau$, for all $e \in E$. }\end{definition}
\vspace{-2mm}

%[Event structure]
\begin{wrapfigure}{r}{.5\textwidth}
\vspace{-10mm}
\begin{center}
  \subfigure[]{\label{fig:event:str:2}\includegraphics[scale = .57]{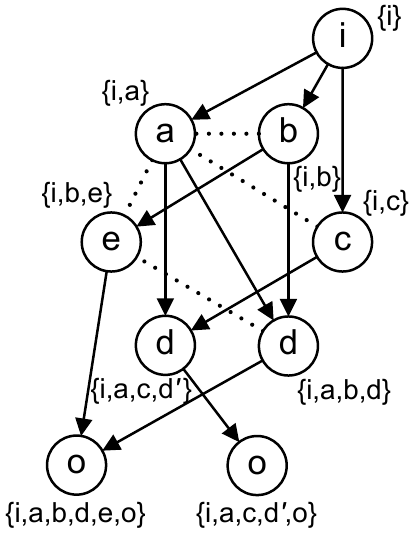}}
  \subfigure[]{\label{fig:event:str:3}\includegraphics[scale = .57, trim = 5mm -3.25mm 0 0]{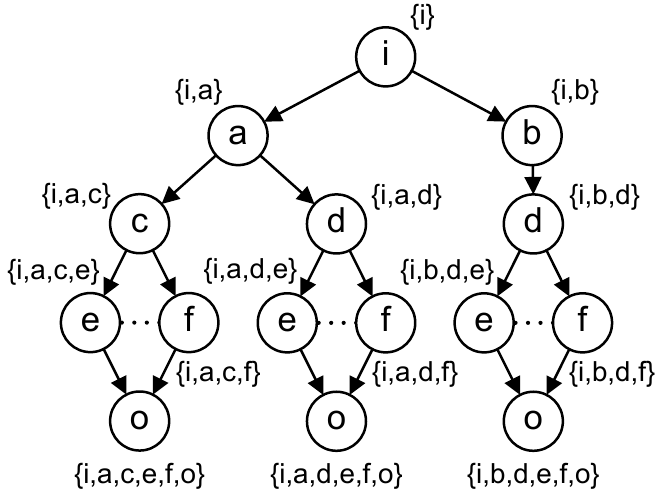}}
\end{center}
\vspace{-7mm}
  \caption{\small{Event structures obtained from (a) \figurename~\ref{fig:running:poset:aug}, and (b) poset of \figurename~\ref{fig:structurable:org}}}
  \label{fig:running:eventstr}
\vspace{-7mm}
\end{wrapfigure}

\noindent \figurename~\ref{fig:event:str:2} visualizes
$\mathcal{P}[\mathcal{L}^\ast[\mathcal{G}]]$ for the graph $\mathcal{G}$ of
\figurename~\ref{fig:running:org}. Events are complete primes of
$\mathcal{L}^\ast[\mathcal{G}]$ (see in boldface in
\figurename~\ref{fig:running:posets} and next to vertices in
\figurename~\ref{fig:running:eventstr}). Directed edges encode causality
(transitive dependencies are not shown), dotted edges represent implicit
concurrency, whereas an absence of an edge hints at a conflict relation. The
event structure in \figurename~\ref{fig:event:str:2} is structurally similar
to the graph in \figurename~\ref{fig:running:org}; they differ only in
relations with fresh $i,o$ events. In general, event structures tend to have
a different structure compared to the originative graphs. For instance,
\figurename~\ref{fig:event:str:3} shows the event structure derived from the
augmented poset of the graph in \figurename~\ref{fig:structurable:org}.

%%%%%%%%%%%%%%%%%%%%%%%%%%%%%%%%%%%%%%%

%\vspace{-1mm}
%\begin{definition}[Def.~18,~\seeCite{NielsenPW81}]\emph{
%\label{def:poset:2:es}
%\newline
%Let $P=(D,\sqsubseteq)$ be a prime algebraic coherent partial order. Then, $\mathcal{P}[P]$ is defined as the event structure $(\mathfrak{P}_P,\leq,\oplus)$, where $\leq$ is $\sqsubseteq$ restricted to $\mathfrak{P}_P$, and for all $e_1,e_2 \in \mathfrak{P}_P : e_1 \oplus e_2$, \ifaof $e_1$ and $e_2$ are inconsistent in $P$.
%}\end{definition}
%\vspace{-1mm}

%\vspace{-1mm}
%\begin{definition}\emph{
%\label{def:es:label}
%Let $\mathcal{G}=(V,A,\mathcal{B},\sigma)$ be an ordering relations graph.
%Let $P=(D,\sqsubseteq)$ be an (augmented) prime algebraic coherent partial order of $\mathcal{G}$.
%Let $\mathcal{E}=(E,\leq,\oplus)$ be $\mathcal{P}[P]$.
%Then, $\lambda : E \rightarrow \mathcal{B}$ is labeling.
%$\lambda(e) = \sigma(v)$, $v \in e \setminus \bigcup_{a \subset e} a$, $e \in E$.
%}\end{definition}
%\vspace{-1mm}

%%%%%%%%%%%%%%%%%%%%%%%%%%%%%%%%%%%%%%%%%%%%%%%%%%%%%%%%%%%%%%%%%%%%%%%%%%%%%%%%%%%%%%%%%%
\vspace{-5mm}
\subsubsection{From event structures to occurrence nets.}
%%%%%%%%%%%%%%%%%%%%%%%%%%%%%%%%%%%%%%%%%%%%%%%%%%%%%%%%%%%%%%%%%%%%%%%%%%%%%%%%%%%%%%%%%%

Nielsen et al. in~\cite{NielsenPW81} show a tight connection between event
structures and occurrence nets. Let $N=(B,E,G)$ be an occurrence net. Then,
$\xi[N] = (E, G^\ast \cap E^2, \#_N \cap E^2)$ is a corresponding event
structure. The next theorem, borrowed from~\cite{NielsenPW81}, defines the
construction of an occurrence net from an event structure.

% Theorem 7
\vspace{-2mm}
\begin{theorem}
\label{thm:7} Let $\mathcal{E}=(E,\leq,\oplus)$, $E \neq \emptyset$, be an
event structure. Then, there exists an occurrence net $\eta[\mathcal{E}]$,
such that $\mathcal{E} = \xi[\eta[\mathcal{E}]]$.
\end{theorem}
\vspace{-5mm}
\begin{proof}
Define the set $\mathit{CE} = \{x \subseteq E \ | \ \forall e_1,e_2 \in x :
e_1 \neq e_2 \Rightarrow e_1 \ \# \ e_2 \}$. The events\pagebreak

\noindent
of $\eta[\mathcal{E}]$ are exactly those in $E$. The set of conditions is
defined by $B = \{\left\langle e,x \right\rangle \ | \ e \in E, x \in
\mathit{CE}, \ and \ \forall e' \in x : e \leq e' \} \cup \{\left\langle 0,x
\right\rangle \ | \ x \in \mathit{CE}, \ and \ x \neq \emptyset \}$. The flow
relation is defined by $G = \{(\left\langle e,x \right\rangle,e') \ | \
\left\langle e,x \right\rangle \in B, e' \in x \} \cup \{(\left\langle 0,x
\right\rangle,e') \ | \ \left\langle 0,x \right\rangle \in B, e' \in x \}
\cup \{(e,\left\langle e,x \right\rangle) \ | \ \left\langle e,x
\right\rangle \in B \}$. It follows, that $\eta[\mathcal{E}]$ is an
occurrence net for which $\# = \oplus$, and hence
$\xi[\eta[\mathcal{E}]]=\mathcal{E}$. \qed
\end{proof}
\vspace{-2mm}

\enlargethispage{\baselineskip}

\noindent \figurename~\ref{fig:running:occurrence_net} shows the occurrence
net which is constructed from the event structure shown in
\figurename~\ref{fig:event:str:2} using the principles of Thm.~\ref{thm:7}.
Thm.~\ref{thm:7} defines a ``maximal'' construction,~\seeCite{NielsenPW81},
\ie the resulting nets tend to contain much redundancy. With
Def.~\ref{def:minimal:occ:net} we aim at preserving only essential behavioral
dependencies.

\begin{figure}[t]
\vspace{-5mm}
\begin{center}
  \includegraphics[width=.9\linewidth]{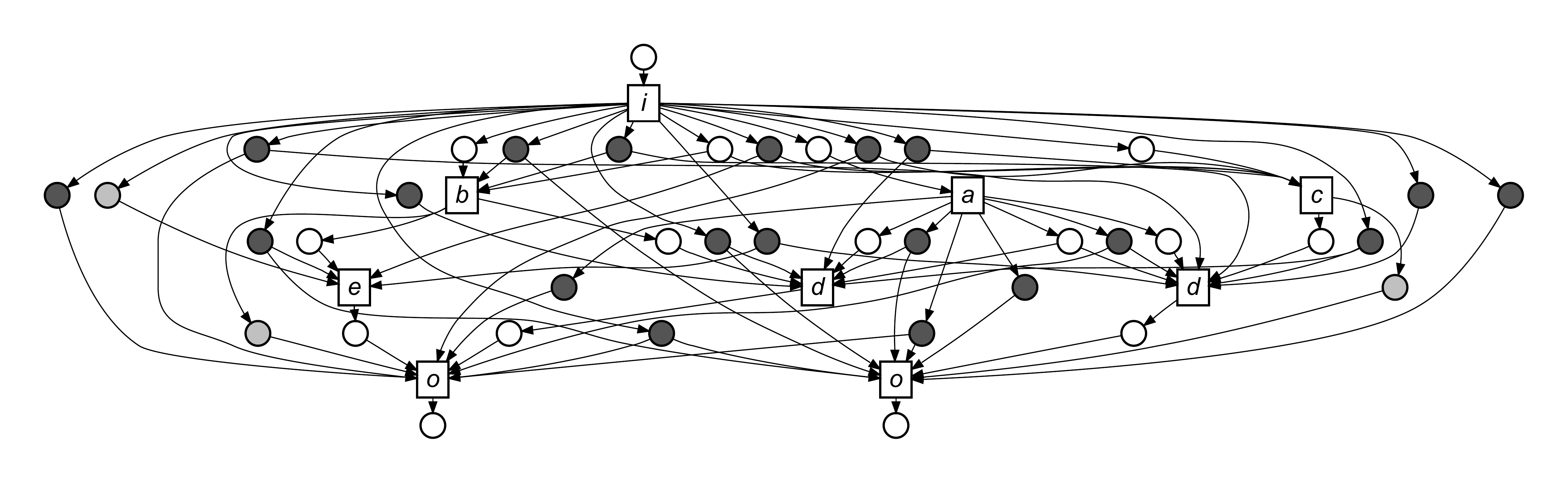}
\end{center}
\vspace{-9mm}
\caption{\small{Occurrence net obtained from Fig.~\ref{fig:event:str:2} by Thm.~\ref{thm:7}.}}
\label{fig:running:occurrence_net}
\vspace{-7mm}
\end{figure}

%\begin{figure}[t]
%\centering
  %\includegraphics[width=\linewidth]{occnet_model12304_all}
  %\caption{\small{Occurrence net obtained from Fig.~\ref{fig:event:str:2} by Thm.~\ref{thm:7}.
  %The net has no redundant conditions; grey-shaded conditions are subsumed or denote transitive conflicts (Def.~\ref{def:minimal:occ:net})
  %}}
%\label{fig:running:occurrence_net}
%\end{figure}

\vspace{-2mm}
\begin{definition}[Conditions]\emph{
\label{def:minimal:occ:net}
%Let $\mathcal{E}=(E,\leq,\oplus)$, $E \neq \emptyset$, be an event structure and
Let $N=(B,E,G)$ be an occurrence net.
\begin{compactitem}
  \item A condition $b \in B$ is \emph{redundant}, \ifaof $\post{b}=\emptyset
\land \exists \ b^\prime \in B, b \neq b^\prime : b^\prime \in
\post{(\pre{b})}$ or $\pre{b}=\emptyset \land \exists \ b^\prime \in B,
b \neq b^\prime : (\post{b} = \post{b^\prime}) \land (\pre{b^\prime}
\neq \emptyset)$.
  \item A condition $b \in B$ is \emph{subsumed} by condition $b^\prime \in
B$, $b \neq b^\prime$, \ifaof $\pre{b}=\pre{b^\prime} \land \post{b}
\subseteq \post{b^\prime}$.
  \item A condition $b \in B$ denotes a \emph{transitive conflict} between
events $e,e^\prime \in E$, \ifaof
$\exists \ b^\prime \in B, b \neq b^\prime \ \exists \ e^{\prime\prime} \in
\post{b^\prime}, e \neq e^{\prime\prime} \neq e^\prime:
\pre{b}=\pre{b^\prime} \land e^\prime \in \post{b} \cap \ \post{b^\prime}
\land \ e \in \post{b} \land \ e \leadsto_N e^{\prime\prime}$.
  \item Any other condition is \emph{required}.
\end{compactitem}
}\end{definition}
\vspace{-2mm}

\begin{wrapfigure}{r}{.37\textwidth}
\vspace{-12mm}
\begin{center}
  \includegraphics[scale = .46]{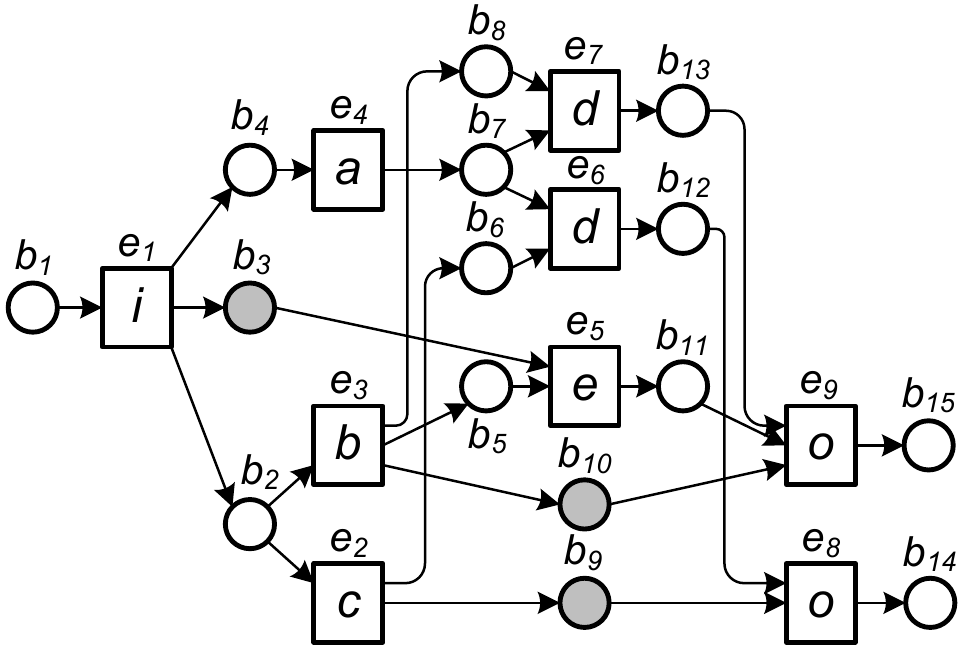}
\end{center}
\vspace{-6mm}
  \caption{\small{Simplified occurrence net obtained from \figurename~\ref{fig:running:occurrence_net}}}
  \label{fig:running:occurrence_net:minimal}
\vspace{-6mm}
\end{wrapfigure}

\noindent
A redundant condition has no pre-event (post-event), and is not a pre-condition (post-condition) of the initial (a final) event. A subsumed
condition $b$ always has a sibling $b'$ expressing the same constraints for
larger set of events; depicted light-grey in
\figurename~\ref{fig:running:occurrence_net}. A condition $b$ denotes a
transitive conflict between two events, if an ``earlier'' condition
$b'$ already denotes this conflict; depicted dark-grey in
\figurename~\ref{fig:running:occurrence_net}. All these conditions can be
removed from the occurrence net without loosing information about ordering of
events.
For our structuring, we remove from an occurrence net all redundant and all
subsumed conditions, and all transitive conflicts which have \emph{at least
two post-events}. Removing these conditions from the net in
\figurename~\ref{fig:running:occurrence_net} yields the net in
\figurename~\ref{fig:running:occurrence_net:minimal}. Note that all
conditions are labeled $\tau$, and that transitive conflicts with one
post-event will be needed for the next step.

%%%%%%%%%%%%%%%%%%%%%%%%%%%%%%%%%%%%%%%

%\begin{theorem}[\theoremname~6, \seeCite{NielsenPW81}]
%\label{thm:6}
%Let $N=(B,E,G)$ be an occurrence net. Then, $\xi[N] = (E, G^\ast \cap E^2, \#_N \cap E^2)$ is an event structure.
%\end{theorem}

%%%%%%%%%%%%%%%%%%%%%%%%%%%%%%%%%%%%%%%%%%%%%%%%%%%%%%%%%%%%%%%%%%%%%%%%%%%%%%%%%%%%%%%%%%
\vspace{-4mm}
\subsubsection{From occurrence nets to nets.}
%%%%%%%%%%%%%%%%%%%%%%%%%%%%%%%%%%%%%%%%%%%%%%%%%%%%%%%%%%%%%%%%%%%%%%%%%%%%%%%%%%%%%%%%%%

The simplified occurrence net obtained by Thm.~\ref{thm:7} and
Def.~\ref{def:minimal:occ:net} is already a process model -- though one with
duplicate structures and multiple sinks. We obtain a more compact model with
a single sink by \emph{folding} the occurrence net.
%
%\begin{wrapfigure}{r}{.37\textwidth}
%\vspace{-12mm}
%\begin{center}
%  \subfigure[]{\label{fig:running:occurrence_net:minimal}\includegraphics[scale = .46, trim = 0 3mm 0 0]{occnet_model12304_2}}
%  \subfigure[]{\label{fig:running:occurrence_net:folded}\includegraphics[scale = .46, trim = 0 3mm 0 7mm]{folded_model12304_2}}
%\end{center}
%\vspace{-7mm}
  %\caption{\small{(a) Simplified occurrence net, and (b) folded net}}
  %\label{fig:running:occnet}
%\vspace{-8mm}
%\end{wrapfigure}
%
%\begin{wrapfigure}[17]{r}{.6\textwidth}\centering
%\vspace{-.8cm}
  %\subfigure[]{\label{fig:running:occurrence_net:minimal}\includegraphics[scale
%= .85]{occnet_model12304}}\hfill
  %\subfigure[]{\label{fig:running:occurrence_net:folded}\includegraphics[scale
%= .85]{folded_model12304}}
%\vspace{-3mm}
  %\caption{\small{Simplified occurrence net and folded net}}
  %\label{fig:running:eventstr}
%\end{wrapfigure}
%
Intuitively, we fold any two nodes of an occurrence

\noindent net which have isomorphic successors into one node. This operation
preserves all ordering relations and all behavior represented in the net.
Folding finite occurrence nets succeeds with the following inductive
definition of a \emph{future equivalence}.

\vspace{-2mm}
\begin{definition}[Future equivalence, Folded net]\emph{
\label{def:future_equivalence}
% Future equivalence
Let $N=(B,E,G,\mathcal{T},\lambda)$ be a labeled occurrence net, $\forall \ b \in B : \lambda(b) = \tau$.
An equivalence relation $\sim_f$ is a \emph{future equivalence} on $N$, \ifaof there exists an equivalence $\mathord{\sim}
\subseteq (B \times B) \cup (E \times E)$:
\begin{compactitem}
  \item For all $b,b^\prime\in B$, if $\post{b}=\post{b^\prime}$, then $b \sim b^\prime$.
  \item For $X,Y \subseteq B \cup E$, write $X \sim Y$, \ifaof $X = \{ x_1,\ldots,x_k\}, Y = \{y_1,\ldots,y_k\}$, s.t.\ $x_i \sim y_i$, for $1
\leq i \leq k$;
  \item For all $x,y \in B \cup E$, if $\lambda(x) = \lambda(y)$ and $\post{x}
\sim \post{y}$ and $\neg ( x \mathrel{||}_N y)$, then $x \sim y$.
\end{compactitem}
The future equivalence defines $x \sim_f y$, \ifaof $x \sim y \wedge (x,y \in E \Rightarrow \pre{x} \sim \pre{y})$.
\newline
% Folded net
Let $\sim_f$ be a future equivalence on $N$; write $\langle
x \rangle_f = \{ y \mid y \sim_f x \}$ for the equivalence class of $x$. Then
the \emph{folded net of $N$ under $\sim_f$} is the net $N_f = ( \{ \langle b
\rangle_f \mid b \in B \}, \{ \langle e \rangle_f \mid e \in E \}, \{
(\langle x \rangle_f, \langle y \rangle_f) \mid (x,y) \in F \}, \lambda_f )$
with $\lambda_f(\langle x \rangle_f) = \lambda(x)$.
}\end{definition}
\vspace{-2mm}

\enlargethispage{\baselineskip}

%\vspace{-1mm}
%\begin{definition}\emph{
%\label{def:folded_net}
%Let $N=(B,E,G,\mathcal{T},\lambda)$ be a labeled occurrence net with $\lambda(b) = \tau$,
%for all $b\in B$. Let $\sim_f$ be a future equivalence on $N$; write $\langle
%x \rangle_f = \{ y \mid y \sim_f x \}$ for the equivalence class of $x$. Then
%the \emph{folded net of $N$ under $\sim_f$} is the net $N_f = ( \{ \langle b
%\rangle_f \mid b \in B \}, \{ \langle e \rangle_f \mid e \in E \}, \{
%(\langle x \rangle_f, \langle y \rangle_f) \mid (x,y) \in F \}, \lambda_f )$
%with $\lambda_f(\langle x \rangle_f) = \lambda(x)$.
%}\end{definition}
%\vspace{-1mm}

\noindent
Considering the occurrence net $N$ in \figurename~\ref{fig:running:occurrence_net:minimal}, the equivalence
$\sim_f$ with the classes $\{b_{14}, b_{15}\}$, $\{e_{8}, e_{9}\}$, $\{b_{9},
b_{11}\}$, $\{b_{12}, b_{13}\}$, $\{e_{6}, e_{7} \}$, $\{b_{6}, b_{8} \}$,
and all other nodes remaining singleton, is a future equivalence on $N$.
Folding $N$ under $\sim_f$ yields the net in \figurename~\ref{fig:running:occurrence_net:folded}.
Folding $N$ into $N_f$ preserves the behavior of $N$,~\see\cite[Thm.~8.7]{FahlandThesis}.

\begin{wrapfigure}{r}{.37\textwidth}
\vspace{-11mm}
\begin{center}
  \includegraphics[scale = .46]{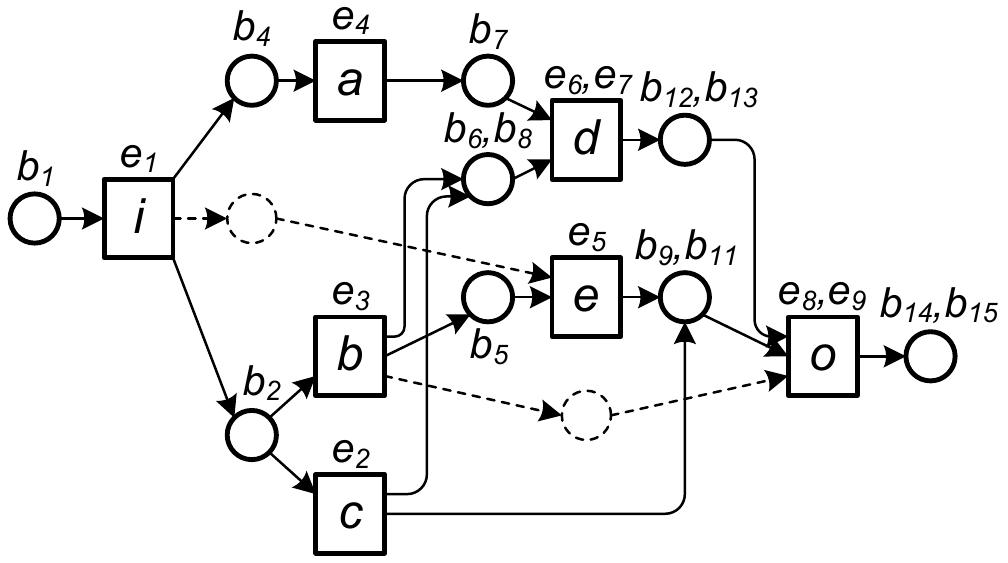}
\end{center}
\vspace{-7mm}
  \caption{\small{Folded net obtained from \figurename~\ref{fig:running:occurrence_net:minimal}}}
  \label{fig:running:occurrence_net:folded}
\vspace{-6mm}
\end{wrapfigure}

Each occurrence net has several future equivalences differing in how
pre-conditions of events are folded. A simple algorithm to compute a future
equivalence implements the steps of Def.~\ref{def:future_equivalence} and
uses branching and backtracking whenever for a condition $b$ there are two or
more pairwise concurrent conditions
%$c_1,c_2,\ldots$
that could be folded
with $b$. Each option is explored and the most-compact folding is
chosen. For instance, after folding $b_{14} \sim_f b_{15}$ and $e_8 \sim_f
e_9$, for $b_{13}$ the folding options $b_{12}$ and $b_{9}$ can be explored;
backtracking yields $b_{12}$ as the better match for $b_{13}$ because of
their $d$-labeled pre-events. Various heuristics improve exploration and
backtracking.

If the original process model has control-flow edges between gateways
\emph{without any visible activity}, folding gets more involved. In this
case, the occurrence net contains supposedly equivalent events with different
numbers of required pre-conditions,~\eg $e_8$ and $e_9$ with required
pre-conditions $\{ b_{12} \}$ and $\{ b_{11}, b_{13} \}$, respectively.
Fortunately, Thm.~\ref{thm:7} encodes all possible invisible control-flow
edges as transitive conflicts with one post-event (grey-shaded conditions in
\figurename~\ref{fig:running:occurrence_net:minimal}). When extending the
future equivalence to pre-conditions of events, a subset of these transitive
conflicts needs to be taken into account as follows:
\begin{compactitem}
\item Pick the largest set $B^\prime$ of required pre-conditions, \eg $b_{11}$ and $b_{13}$.
\item For each $b \in B^\prime$, extend the folding equivalence with a required condition or a transitive conflicts, \eg $b_{13} \sim_f
b_{12}$, $b_{11} \sim_f b_9$.
\item Finally, remove all transitive conflicts not required in this step, \eg $b_{10}$.
\end{compactitem}
Applying this procedure on our example yields the folded net shown in
\figurename~\ref{fig:running:occurrence_net:folded} without the dashed
conditions and arcs.

%%%%%%%%%%%%%%%%%%%%%%%%%%%%%%%%%%%%%%%%%%%%%%%%%%%%%%%%%%%%%%%%%%%%%%%%%%%%%%%%%%%%%%%%%%
\vspace{-4mm}
\subsubsection{From nets to process models.}
%%%%%%%%%%%%%%%%%%%%%%%%%%%%%%%%%%%%%%%%%%%%%%%%%%%%%%%%%%%%%%%%%%%%%%%%%%%%%%%%%%%%%%%%%%

The folding was the second to last step in synthesizing a process model from
a given ordering relations graph. We obtained
a Petri net $N_f$ which we now transform into a process model $P$.

The initial transition $i$ (final transition $o$) is mapped to the start
(end) node of $P$. Every other transition of $N_f$ becomes a task of $P$.
Gateways of $P$ follow 

\begin{wrapfigure}{r}{.37\textwidth}
\vspace{-3mm}
\begin{center}
  \includegraphics[scale = .5]{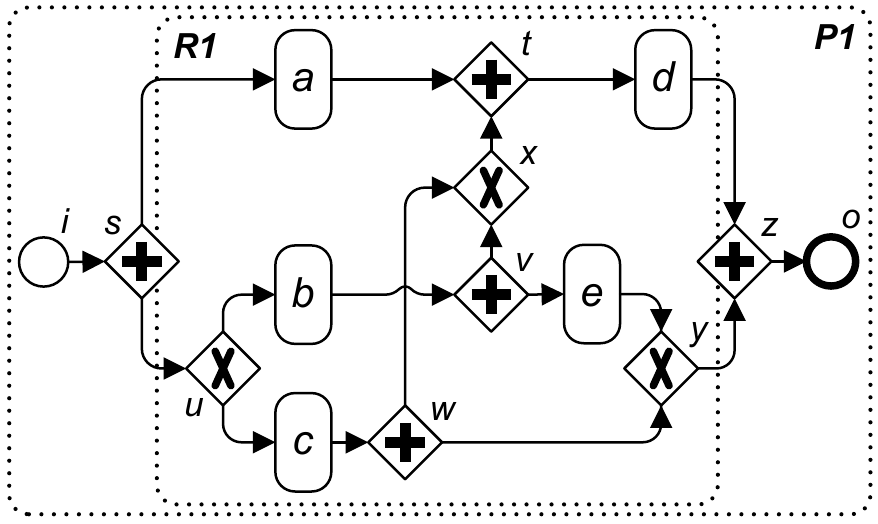}
\end{center}
\vspace{-6.5mm}
\caption{\small{Process model obtain\-ed from \figurename~\ref{fig:running:occurrence_net:folded}}}
\label{fig:total:prim}
\vspace{-6.5mm}
\end{wrapfigure}

\noindent
from non-singleton pre- and postsets of nodes of
$N_f$. A transition $t$ with two or more pre-places is preceded by an $and$
join; two or more post-places of $t$ define an $and$ split; the pre- and
postsets of places define $xor$ splits and joins, respectively; $and$
gateways are always positioned closer to the task. In our example,
$\post{e_1}$ defines $and$ split $s$ in \figurename~\ref{fig:total:prim},
$\post{b_2}$ defines $xor$ split $u$, $\post{e_3}$ defines $and$ split $v$,
$\pre{ \langle e_6,e_7 \rangle }$ defines $and$ join $t$, and $\pre{ \langle
b_6,b_8 \rangle }$ defines $xor$ join $x$ positioned between $t$ and $v$
($and$ gateways closer to tasks); correspondingly for all other gateways. The
arc from $e_2$ to $\langle b_{9},b_{11} \rangle$ which was obtained from a
transitive conflict (Def.~\ref{def:minimal:occ:net}) results in an important
control-flow arc from $w$ to $y$ without any task.

%% file: conclusion.tex
\vspace{-2.6mm}
\section{Related Work and Conclusion}
\label{sec:conclusion}
\vspace{-1.75mm}
%%%%%%%%%%%%%%%%%%%%%%%%%%%%%%%%%%%%%%%%%%%%%%%%%%%%%%%%%%%%%%%%%%%%%%%%%%%%%%%%%%%%%%%%%%

%\begin{compactitem}
%  \item Can be used for automated generation of process models, but one needs to characterize correct (sound) graphs.
%  \item Can be done more efficiently. Our aim was to be close to existing theory.
%  \item Implemented in BPStruct
%\end{compactitem}

%%%%%%%%%%%%%%%%%%%%%%%%%%%%%%%%%%%%%%%%%%%%%%%%%%%%%%%%%%%%%%%%%%%%%%%%%%%%%%%%%%%%%%%%%%

\enlargethispage{\baselineskip}

In this paper, we addressed the problem of structuring acyclic process
models. It is well known that any flowchart can be structured~\cite{O82}, but
the same claim does not apply for process models comprising
concurrency~\cite{KHB00}. Some works have been devoted to the
characterization of sources of unstructuredness~\cite{LK05,PGW09} and to
development of methods for structuring process models with
concurrency~\cite{HK04,HFKV08}. In~\cite{PGD10}, we presented the first full
characterization of the class of acyclic process models that have an
equivalent structured version along with a structuring method. The method
stops when the input model contains an inherently unstructured fragment. This
paper completes the approach by providing a method to synthesize the
fragments corresponding to inherently unstructured parts of the input model.

\vspace{-0.1mm}

Close to our setting, the problem of synthesizing nets from behavioral
specifications has been a line of active research for about two decades~\cite{CortadellaKLY98,BergenthumDM09}. This area has given rise to a
rich body of knowledge and to a number of tools, \eg {\ttfamily
petrify}~\cite{CortadellaKLY98} and {\ttfamily viptool}~\cite{BergenthumDM09}. 
Yet, these solutions fail in our setting:
{\ttfamily petrify} aims at maximizing concurrency while our synthesis
preserves given concurrency, {\ttfamily viptool} synthesizes nets with
arc weights, which do not map to process models.

\vspace{-0.1mm}

The approach is implemented in a tool, namely {\ttfamily bpstruct}, which is
publicly available at \url{http://code.google.com/p/bpstruct}. The running
time of our structuring technique is mostly dominated by the time required to
compute proper prefixes, which for safe systems is $O((|B| /
n)^n)$~\cite{EsparzaRV02}, where $B$ is the set of conditions of the prefix
and $n$ is the maximal size of the presets of the transitions in the
originative system. All other steps can be accomplished in linear time.
Concerning the extension for maximal structuring, the theoretic discussion in
this paper implies exponential time and space complexity when constructing
posets (this is due to our wish to be close to the existing theory). However,
in practice, given an ordering relations graph one can construct a poset
which only contains information from the graph, without introducing duplicate
events, and thus stay linear to the size of the graph. At the theoretical
level this requires introduction of a concept of a cutoff for posets followed
by an adjustment of the theories along subsequent transformation steps. The
folding step is a reverse of unfolding and, thus, in the best case can be
performed in the same time. The fact that the running time depends on the
size of the result, allows introduction 

\noindent
of a heuristic to terminate
computation if the result gets large, \eg the event duplication factor is
larger than two. However, in practice we have never observed such a need with
our implementation always delivering the result in milliseconds. Our ongoing
work aims at extending the method to handle models with loops.

%Our ongoing work aims at extending the method to handle models containing loops with arbitrary topologies comprising complex overlapping networks of $and$ and $xor$ gateways. We also aim at completing the coverage of missing modeling constructs found in BPMN and similar notations, such as $or$ gateways, complex gateways, exception handlers and non-interrupting events. The characterization of ordering relations graphs as a kind of event structure and their relation with other higher order event structures, \eg flow event structures~\cite{Boudol90}, are part of our research agenda.

\enlargethispage{\baselineskip}